\documentclass[journal]{IEEEtran}
\usepackage{cite,amsmath}
\usepackage{subfig}
\usepackage{color}
\usepackage{graphicx}
\usepackage{pstool}

\newcommand{\ve}[1]{\boldsymbol{#1}}
\newcommand{\te}[1]{\overline{\overline{#1}}}

\newcounter{tempEquationCounter}
\newcounter{thisEquationNumber}
\newenvironment{floatEq}
{\setcounter{thisEquationNumber}{\value{equation}}\addtocounter{equation}{1}
\begin{figure*}[!t]
\normalsize\setcounter{tempEquationCounter}{\value{equation}}
\setcounter{equation}{\value{thisEquationNumber}}
}
{\setcounter{equation}{\value{tempEquationCounter}}
\hrulefill\vspace*{4pt}
\end{figure*}
}

\begin{document}

\title{Homogenization and Scattering Analysis of Second-Harmonic Generation in\\ Nonlinear Metasurfaces}

\author{Karim~Achouri, Gabriel D. Bernasconi, J\'{e}r\'{e}my Butet, and Olivier J. F. Martin

}

\maketitle

\begin{abstract}
We propose an extensive discussion on the homogenization and scattering analysis of second-order nonlinear metasurfaces. Our developments are based on the generalized sheet transition conditions (GSTCs) which are used to model the electromagnetic responses of nonlinear metasurfaces. The GSTCs are solved both in the frequency domain, assuming an undepleted pump regime, and in the time-domain, assuming dispersionless material properties but a possible depleted pump regime. Based on these two modeling approaches, we derive the general second-harmonic reflectionless and transmissionless conditions as well as the conditions of asymmetric reflection and transmission. We also discuss and clarify the concept of nonreciprocal scattering pertaining to nonlinear metasurfaces.
\end{abstract}


\IEEEpeerreviewmaketitle


\section{Introduction}

Over the past decade, metasurfaces have attracted tremendous attention due to their incredible capabilities to control electromagnetic waves together with their low weight, reduced thickness and relative ease of fabrication~\cite{capasso1,yu2014flat,Minovich2015,Glybovski20161,Kildishev1232009}. More recently, nonlinear metasurfaces made their apparition and were rapidly considered as a new paradigm shift in nonlinear optics~\cite{Alu2017Paradigm,Alu2014Giant}. Indeed, the capability of engineering the shape and the composition of the nonlinear scattering particles composing the metasurfaces allows one to strongly confine the electromagnetic fields and thus achieve very high nonlinear effects, which may be several orders of magnitude stronger than those obtainable in conventional nonlinear crystals~\cite{Alu2014Giant,yang2015nonlinear,tymchenko2017THz}. Additionally, nonlinear metasurfaces may be used to manipulate the second- or third-harmonic scattered fields in a very elaborate fashion and with much more degrees of freedom compared to our current ability of controlling linear fields in conventional linear metasurfaces~\cite{Kuittinen2013SGH,Celebrano2015,KuangYu2017WaveVector,Zentgraf2017}. For instance, it was suggested that nonlinear metasurfaces may be used to realize functionalities such as optical switches, diodes and transistors~\cite{chen2010optical}, and also generate multiple holographic images based on the Pancharatnam-Berry phase effect~\cite{Ye2016,Alu2016Pancharatnam}. We also note the particularly interesting potential of metasurfaces to exhibit magnetic nonlinear effects~\cite{Kruk2017NLmagnetic,Kruk2016VectorBeams,Kruk2015EnhancedMagnetic}. This ability to control both electric and magnetic nonlinear responses is of utmost importance for a complete control of the nonlinear scattered fields, as will be discussed thereafter.

Even though nonlinear metasurfaces are already quite attractive, there still needs to be major theoretical and practical works to be done before they achieve their full potential. This paper tackles the theoretical aspect of second-harmonic generation in second-order nonlinear metasurfaces. The mathematical basis of this work is a direct extension of our previous works on linear~\cite{achouri2014general,Achouri2015c,achouri2017design} and nonlinear metasurfaces~\cite{Achouri2017NLMS}. It also extends other similar theoretical works pertaining to second-order nonlinear metasurfaces~\cite{Smith2012NL,Smith2013NLuniderc,poutrina2014multipole,Poutrina2016,LIU201853}. The theory presented in these papers mostly refers to relatively specific situations, while we try here to be as general as possible in our description of second-harmonic scattering.

The goal of this paper is to provide the reader with a comprehensive discussion on the homogenization and scattered field analysis of second-order nonlinear metasurfaces and more specifically on their properties in terms of second-harmonic generation. Accordingly, we propose a frequency- and time-domain analysis of nonlinear metasurfaces, upon which we notably derive the general reflectionless and transmissionless second-harmonic conditions. The mathematical model that we use is based on rigorous zero-thickness transition conditions, which have been extensively used in the case of linear metasurfaces~\cite{kuester2003av,holloway2012overview,achouri2014general,Achouri2015c,achouri2017design}. This model relates the incident, reflected and transmitted fields to the metasurface linear and now also nonlinear susceptibilities. Finally, we also propose a discussion on the nonreciprocal behavior of nonlinear metasurfaces and, more generally, nonlinear optical structures~\cite{potton2004reciprocity,Hubner00TRSinNL,Zheng2013Timereversed,naguleswaran1998onsager}. This is to clarify certain misconceptions about the definition of nonreciprocity in nonlinear optics.

This paper is organized as follows: In Sec.~\ref{Sec:mod}, we start by discussing the electromagnetic modeling of nonlinear metasurfaces. More specifically, we explain how these metasurfaces may be mathematically modeled by zero-thickness transition conditions. Then, In Secs.~\ref{Sec:freq} and~\ref{Sec:time}, we respectively present a frequency-domain and a time-domain approach to perform the homogenization and the scattering analysis of these metasurfaces. In Sec.~\ref{Sec:conc}, we briefly conclude our discussion. Additionally, we also provide several useful considerations in the appendices. These include a discussion on the homogenization and scattering analysis of linear metasurfaces in Appendix~\ref{app:Sparam}, and a discussion which aims at clarifying the nonreciprocal behavior of linear and nonlinear metasurfaces in Appendix~\ref{app:recip}. Finally, we provide the general reflectionless conditions of linear metasurfaces in Appendix~\ref{app:RLC}.

\section{GSTCs Modeling of Second-Order\\ Nonlinear Metasurfaces}
\label{Sec:mod}

\begin{figure}[h!]
\centering
\subfloat[]{\label{fig:MSfig1}
\includegraphics[width=0.8\columnwidth]{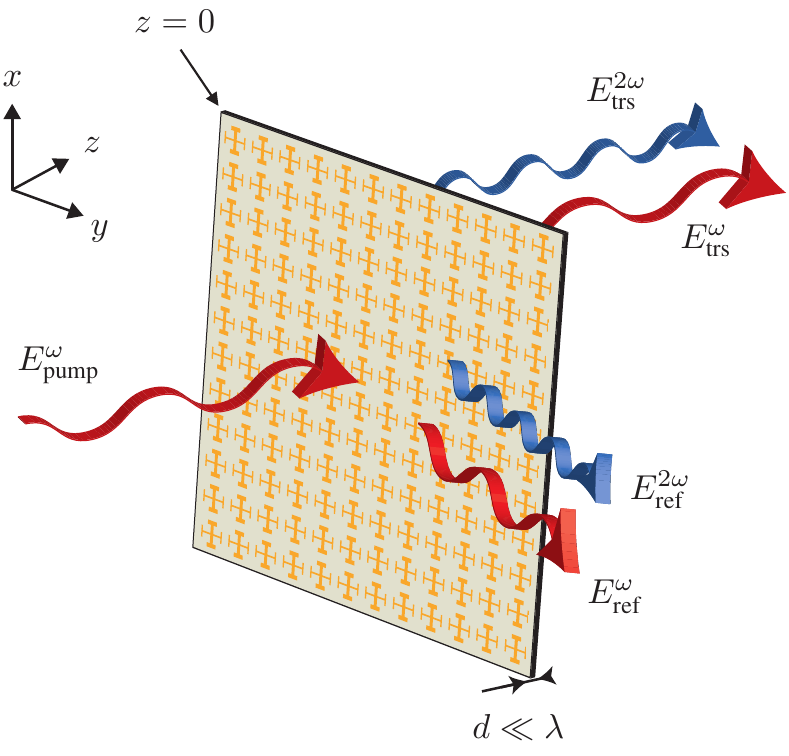}
}\\
\subfloat[]{\label{fig:MSfig2}
\includegraphics[width=0.8\columnwidth]{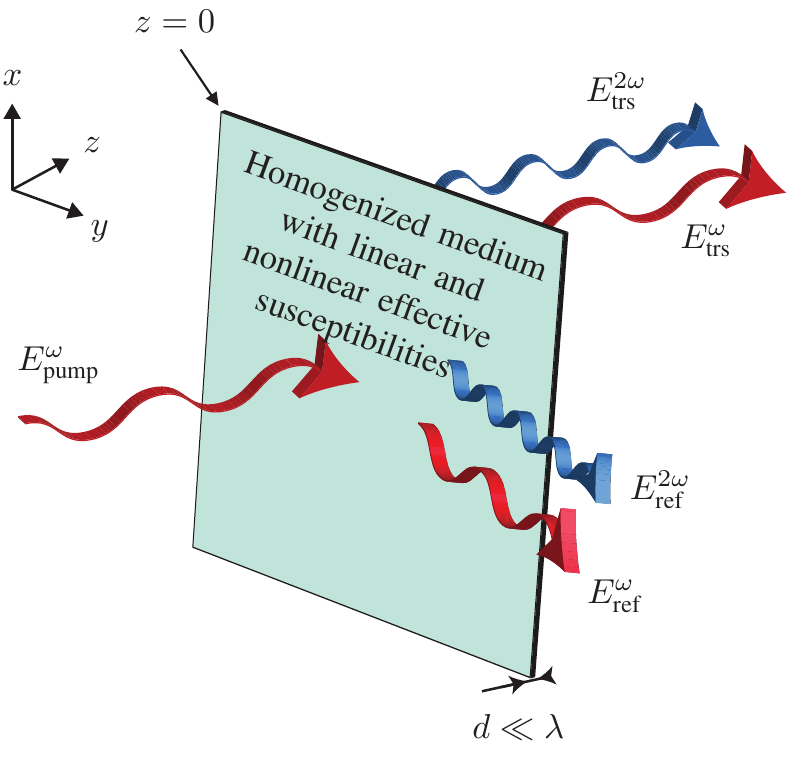}
}
\caption{Pump field at frequency $\omega$ being scattered at frequencies $\omega$ and $2\omega$ due to its interactions with second-order periodically uniform nonlinear metasurfaces. (a) Physical metasurface built from nonlinear scattering particles. (b) Homogenized metasurface with effective susceptibilities that produce the same scattering as that in~(a).}
\label{fig:MSfig}
\end{figure}

Let us consider a periodically uniform metasurface made of second-order nonlinear scattering particles lying in the $xy$-plane at $z=0$, as illustrated in Fig.~\ref{fig:MSfig1}. The period of the scattering particles within the plane of the metasurface is also much smaller than the wavelength with a typical unit cell lateral dimension of $\lambda/3$ to $\lambda/5$. From the perspective of an incident pump field at frequency $\omega$, the metasurface appears as a perfectly uniform, homogeneous and time invariant medium with effective susceptibilities independent of the coordinates $(x,y,t)$, as suggested in Fig.~\ref{fig:MSfig2}. Note that, in these figures, we only represent the scattered fields at frequencies $\omega$ and $2\omega$ for convenience, while there may be higher-order harmonics generated by the metasurface.

Metasurfaces may, in most cases, be conveniently modeled assuming that they consist of zero-thickness sheets of excitable electric and magnetic polarizations~\cite{kuester2003av,Holloway2009}. This assumption of zero-thickness is a very good approximation since metasurfaces are electrically thin structures with a typical thickness, $d$, much smaller than the operation wavelength ($d \ll \lambda$). When a metasurface is modeled as a zero-thickness sheet, it is possible to relate its interactions with electromagnetic fields in a general and complete fashion using \emph{generalized sheet transition conditions} (GSTCs)~\cite{Idemen1973,kuester2003av,achouri2014general}.

From a general perspective, the relationship between the incident, reflected and transmitted fields and the linear and nonlinear susceptibilities of the metasurface may be expressed through the GSTCs. In the time-domain, assuming a time dependence $e^{j\omega t}$, the GSTCs are given by~\cite{Idemen1973}
\begin{subequations}
\label{eq:GSTCs}
\begin{equation}
\ve{z}\times\Delta\ve{H}=\frac{\partial}{\partial t}\ve{P} - \ve{z}\times \nabla M_z,
\end{equation}
\begin{equation}
\ve{z}\times\Delta\ve{E}=-\mu_0\frac{\partial}{\partial t}\ve{M} - \frac{1}{\epsilon_0}\ve{z}\times \nabla P_z,
\end{equation}
\end{subequations}
where $\ve{P}$ and $\ve{M}$ are the metasurface electric and magnetic polarization densities, proportional to the susceptibilities, and $\Delta\ve{E}$ and $\Delta\ve{H}$ are the differences of the electric and magnetic fields on both sides of the metasurface, respectively~\cite{achouri2014general}.

The GSTCs may be used to perform three distinct operations~\cite{achouri2014general,Achouri2015c,achouri2017design}. They may be used to a)~\emph{synthesize} -- find the susceptibility functions -- of a metasurface so that it scatters light in a specified fashion. In this case, the susceptibilities are mathematical functions and the corresponding scattering particles need to be practically designed and realized. The GSTCs may also be used to b)~\emph{homogenize} -- find the effective linear and nonlinear susceptibilities of -- a metasurface made of scattering particles with specific shapes and materials. Finally, they may be used to c)~\emph{analyze} -- find the fields scattered by -- a metasurface with known (effective) susceptibilities. These operations may be realized by solving~\eqref{eq:GSTCs} to either express the susceptibilities in terms of known electromagnetic fields (specified for the synthesis, and simulated or measured for the homogenization) or to express the scattered fields in terms of known susceptibilities to analyze the metasurface scattering.

In this work, we will mostly discuss the last two operations, i.e. the homogenization and scattering analysis of second-order nonlinear metasurfaces. This is because the synthesis of such structures, although possible, is generally difficult due to the very large number of susceptibility components, as will be discussed shortly. Moreover, we will concentrate our attention on the generation of second harmonic light since the study of linear scattering is already very well documented in the literature. We will also neglect the presence of all higher-order harmonics that may emerge from the interactions of waves at frequencies $\omega$ and $2\omega$ (and so on) with the nonlinear medium.

In order to solve the system~\eqref{eq:GSTCs}, we will consider two different approaches. A frequency-domain approach, for which we will assume that the pump at frequency $\omega$ is undepleted and that the susceptibilities are dispersive so that in general $\chi(\omega) \neq \chi(2\omega)$, and a time-domain approach, for which the pump may be depleted but the susceptibilities are not dispersive so that $\chi(\omega) = \chi(2\omega)$. As we will see, these two approaches present their own benefits and drawbacks. Therefore, the most adequate approach mostly depends on the needs of the user.

Note that the fact that the metasurface is spatially uniform implies that no diffraction orders will be produced upon light scattering. Therefore, an incident plane wave, impinging on the metasurface at arbitrary angles $(\theta_\text{i}, \phi_\text{i})$, will be reflected and transmitted with reflection and transmission angles obeying Snell's law. This is of particular importance since, for any incidence angles, we know how the metasurface will scatter light, which greatly simplifies the operations of homogenization and scattered field analysis.

\section{Frequency-Domain Approach}
\label{Sec:freq}

In this frequency-domain approach, we consider by approximation that the interactions of an incident wave at frequency $\omega$ with the metasurface linear and nonlinear susceptibilities produce scattered fields \emph{only} at frequencies $\omega$ and $2\omega$, respectively. This assumption is almost always valid since the amplitudes of the higher-order ($>$3) harmonics generated by a second-order nonlinear material are negligible~\cite{boyd2003nonlinear}. We also consider that the pump is not depleted. This assumption violates conservation of energy but it greatly simplifies the modeling of nonlinear metasurfaces and still provides excellent results as long as the power of the second-harmonic is much smaller than that of the pump, which is usually the case with current nonlinear metasurfaces~\cite{LIU201853}.

With this assumption of undepleted pump, it is trivial to homogenize and analyze the response of metasurfaces at frequency $\omega$. It may be easily done following well known techniques applied to linear bianisotropic metasurfaces. In Appendix~\ref{app:Sparam}, we briefly summarize the main steps to obtain the effective \emph{linear} susceptibilities of a metasurface and to compute the fields scattered by a metasurface with known effective susceptibilities.

We are now interested in homogenizing the \emph{nonlinear} susceptibilities and analyzing the second-harmonic scattered waves of a second-order nonlinear metasurface. To do so, we consider the GSTCs in~\eqref{eq:GSTCs} and write them in the frequency-domain so that they relate the scattered fields and the polarization densities both at frequency $2\omega$. They thus read\footnote{Note that the time derivatives in~\eqref{eq:GSTCs} reduce to $j2\omega$ in this frequency-domain representation.}
\begin{subequations}
\label{eq:freqGSCTs}
\begin{equation}
\ve{z}\times\Delta\ve{H}^{2\omega}=j2\omega\ve{P}^{2\omega} - \ve{z}\times \nabla M_z^{2\omega},
\end{equation}
\begin{equation}
\ve{z}\times\Delta\ve{E}^{2\omega}=-j2\omega\mu_0\ve{M}^{2\omega} - \frac{1}{\epsilon_0}\ve{z}\times \nabla P_z^{2\omega},
\end{equation}
\end{subequations}
where the electric and magnetic polarization densities are split into linear and nonlinear terms as
\begin{subequations}
\label{eq:SplitPM}
\begin{equation}
\ve{P}^{2\omega} = \ve{P}_{\text{lin}}^{2\omega} + \ve{P}_{\text{nl}}^{2\omega},
\end{equation}
\begin{equation}
\ve{M}^{2\omega} = \ve{M}_{\text{lin}}^{2\omega} + \ve{M}_{\text{nl}}^{2\omega}.
\end{equation}
\end{subequations}
The linear polarizations correspond to the interactions of the fields at $2\omega$ with the linear metasurface susceptibilities at that frequency and may be generally expressed as
\begin{subequations}
\label{eq:PM}
\begin{equation}
\ve{P}_{\text{lin}}^{2\omega} =\epsilon_0 \te{\chi}_\text{ee}\cdot\ve{E}_\text{av}^{2\omega} + \epsilon_0\eta_0 \te{\chi}_\text{em}\cdot\ve{H}_\text{av}^{2\omega},
\end{equation}
\begin{equation}
\ve{M}_{\text{lin}}^{2\omega}= \te{\chi}_\text{mm}\cdot\ve{H}_\text{av}^{2\omega} + \frac{1}{\eta_0} \te{\chi}_\text{me}\cdot\ve{E}_\text{av}^{2\omega},
\end{equation}
\end{subequations}
where $\eta_0$ is the vacuum impedance, $\ve{E}_\text{av}^{2\omega}$ and $\ve{H}_\text{av}^{2\omega}$ are the arithmetic averages of the fields on both sides of the metasurface, and $\te{\chi}_\text{ee}, \te{\chi}_\text{mm}, \te{\chi}_\text{me}$ and $\te{\chi}_\text{em}$ are the electric, magnetic, magnetoelectric and electromagnetic linear susceptibility tensors, respectively~\cite{lindell1994electromagnetic}.

The nonlinear polarizations correspond to the interactions of the pump fields at $\omega$ with the nonlinear metasurface susceptibilities. These polarizations can thus be considered as the sources of the second-harmonic generation. Taking into account all possible interactions of the fields, they may be generally expressed as
\begin{subequations}
\label{eq:NLPM}
\begin{equation}
\begin{split}
\ve{P}_{\text{nl}}^{2\omega} =& \frac{1}{2}\epsilon_0\big(\te{\chi}_\text{eee}:\ve{E}_\text{av}^\omega\ve{E}_\text{av}^\omega + \eta_0\te{\chi}_\text{eem}:\ve{E}_\text{av}^\omega\ve{H}_\text{av}^\omega \\
&+ \eta_0^2\te{\chi}_\text{emm}:\ve{H}_\text{av}^\omega\ve{H}_\text{av}^\omega \big),
\end{split}
\end{equation}
\begin{equation}
\begin{split}
\ve{M}_{\text{nl}}^{2\omega} =& \frac{1}{2}\big(\eta_0\te{\chi}_\text{mmm}:\ve{H}_\text{av}^\omega\ve{H}_\text{av}^\omega + \te{\chi}_\text{mem}:\ve{E}_\text{av}^\omega\ve{H}_\text{av}^\omega \\
&+ \frac{1}{\eta_0}\te{\chi}_\text{mee}:\ve{E}_\text{av}^\omega\ve{E}_\text{av}^\omega\big),
\end{split}
\end{equation}
\end{subequations}
where we have used our own convention for the dimension of the susceptibilities, by introducing the impedance parameter, such that they all have the same dimension. Note that the factor $\frac{1}{2}$ in front of relations~\eqref{eq:NLPM} comes from the time derivatives in~\eqref{eq:GSTCs}. This factor appears in~\eqref{eq:NLPM} and not in~\eqref{eq:PM} is because
\begin{equation}
\label{eq:fac1}
\frac{\partial}{\partial t}\ve{E}_\text{av}^{2\omega}~ \propto~ \frac{\partial}{\partial t} \cos{(2\omega t)} = - 2\omega \sin{(2\omega t)},
\end{equation}
while
\begin{equation}
\label{eq:fac2}
\frac{\partial}{\partial t}\ve{E}_\text{av}^{\omega}\ve{E}_\text{av}^{\omega}~ \propto~ \frac{\partial}{\partial t} \cos^2{(\omega t)} = - \omega \sin{(2\omega t)}.
\end{equation}
Therefore, a factor of $\frac{1}{2}$ must be considered for the nonlinear polarization densities.

\subsection{Homogenization Technique}
\label{sec:homotech}

In order to homogenize and thus find the effective nonlinear susceptibilities of a metasurface, we have to solve~\eqref{eq:freqGSCTs} along with relations~\eqref{eq:SplitPM} to~\eqref{eq:NLPM}. Considering plane wave illumination, the general nonlinear GSTCs thus form the following set of 4 equations, which are given in reduced tensor notation for convenience\footnote{Note that in these expressions, we have dropped the summation terms such that: $A^{ij}B^{ij}=\sum_{i,j} A^{ij}B^{ij}$. We have also dropped the ``av'' term for the fields multiplying the susceptibilities for conciseness.}:
\begin{subequations}
\label{eq:homo}
\begin{equation}
\begin{split}
&-\Delta H^{2\omega}_y = j2\omega\epsilon_0(\chi_\text{ee}^{xi}E_i^{2\omega}+\eta_0\chi_\text{em}^{xi}H_i^{2\omega} +\frac{1}{2}\chi_\text{eee}^{xij}E_i^\omega E_j^\omega \\
&\quad+\frac{1}{2}\eta_0\chi_\text{eem}^{xij}E_i^\omega H_j^\omega +\frac{1}{2}\eta_0^2\chi_\text{emm}^{xij}H_i^\omega H_j^\omega)-jk_y^{2\omega}(\chi_\text{mm}^{zi}H_i^{2\omega}\\
&\quad+\frac{1}{\eta_0}\chi_\text{me}^{zi}E_i^{2\omega} +\frac{1}{2}\eta_0\chi_\text{mmm}^{zij}H_i^\omega H_j^\omega +\frac{1}{2}\chi_\text{mem}^{zij}E_i^\omega H_j^\omega \\
&\quad +\frac{1}{2\eta_0}\chi_\text{mee}^{zij}E_i^\omega E_j^\omega),
\end{split}
\end{equation}
\begin{equation}
\begin{split}
&\Delta H^{2\omega}_x = j2\omega\epsilon_0(\chi_\text{ee}^{yi}E_i^{2\omega}+\eta_0\chi_\text{em}^{yi}H_i^{2\omega} +\frac{1}{2}\chi_\text{eee}^{yij}E_i^\omega E_j^\omega\\
&\quad +\frac{1}{2}\eta_0\chi_\text{eem}^{yij}E_i^\omega H_j^\omega +\frac{1}{2}\eta_0^2\chi_\text{emm}^{yij}H_i^\omega H_j^\omega)+jk_x^{2\omega}(\chi_\text{mm}^{zi}H_i^{2\omega}\\
&\quad+\frac{1}{\eta_0}\chi_\text{me}^{zi}E_i^{2\omega} +\frac{1}{2}\eta_0\chi_\text{mmm}^{zij}H_i^\omega H_j^\omega +\frac{1}{2}\chi_\text{mem}^{zij}E_i^\omega H_j^\omega \\
&\quad +\frac{1}{2\eta_0}\chi_\text{mee}^{zij}E_i^\omega E_j^\omega),
\end{split}
\end{equation}
\begin{equation}
\begin{split}
&-\Delta E^{2\omega}_y = -j2\omega\mu_0(\chi_\text{mm}^{xi}H_i^{2\omega}+\frac{1}{\eta_0}\chi_\text{me}^{xi}E_i^{2\omega} \\
&\quad+\frac{1}{2}\eta_0\chi_\text{mmm}^{xij}H_i^\omega H_j^\omega +\frac{1}{2}\chi_\text{mem}^{xij}E_i^\omega H_j^\omega +\frac{1}{2\eta_0}\chi_\text{mee}^{xij}E_i^\omega E_j^\omega)\\
&\quad-jk_y^{2\omega}(\chi_\text{ee}^{zi}E_i^{2\omega}+\eta_0\chi_\text{em}^{zi}H_i^{2\omega} +\frac{1}{2}\chi_\text{eee}^{zij}E_i^\omega E_j^\omega \\
&\quad +\frac{1}{2}\eta_0\chi_\text{eem}^{zij}E_i^\omega H_j^\omega +\frac{1}{2}\eta_0^2\chi_\text{emm}^{zij}H_i^\omega H_j^\omega),
\end{split}
\end{equation}
\begin{equation}
\begin{split}
&\Delta E^{2\omega}_x = -j2\omega\mu_0(\chi_\text{mm}^{yi}H_i^{2\omega}+\frac{1}{\eta_0}\chi_\text{me}^{yi}E_i^{2\omega} +\frac{1}{2}\eta_0\chi_\text{mmm}^{yij}H_i^\omega H_j^\omega \\
&\quad+\frac{1}{2}\chi_\text{mem}^{yij}E_i^\omega H_j^\omega +\frac{1}{2\eta_0}\chi_\text{mee}^{yij}E_i^\omega E_j^\omega)+jk_x^{2\omega}(\chi_\text{ee}^{zi}E_i^{2\omega}\\
&\quad+\eta_0\chi_\text{em}^{zi}H_i^{2\omega} +\frac{1}{2}\chi_\text{eee}^{zij}E_i^\omega E_j^\omega +\frac{1}{2}\eta_0\chi_\text{eem}^{zij}E_i^\omega H_j^\omega \\
&\quad+\frac{1}{2}\eta_0^2\chi_\text{emm}^{zij}H_i^\omega H_j^\omega),
\end{split}
\end{equation}
\end{subequations}
where $i,j = \{x,y,z\}$ and where we have transformed the gradients in~\eqref{eq:freqGSCTs} into $\nabla \rightarrow -jk_u^{2\omega}$ with $u=\{x,y\}$ since we are considering plane waves and where $j = \sqrt{-1}$ is not to be confused with the tensor notation. Note that the $\frac{1}{2}$ factor discussed in Eqs.~\eqref{eq:fac1} and~\eqref{eq:fac2} must also exists in the case of the spatial derivatives in~\eqref{eq:freqGSCTs}.

As discussed in Appendix~\ref{app:Sparam}, the linear susceptibilities in~\eqref{eq:homo} can be obtained using the retrieval method applied to linear metasurfaces~\cite{LIU201853}. For the nonlinear susceptibilities, the system~\eqref{eq:homo} must be solved so as to express the susceptibilities in terms of the incident, reflected and transmitted fields at frequency $2\omega$ that may be found from numerical simulations. However, it is clear that this system of equations is, in most cases, largely under-determined. Indeed, assuming that the nonlinear susceptibilities only have intrinsic permutation symmetry ($\chi^{ijk} = \chi^{ikj}$ as is the case of most nonlinear systems~\cite{boyd2003nonlinear}) there is a total number of 108 unknown susceptibility components in~\eqref{eq:homo} for only 4 equations. Note that, in the case of linear susceptibilities, some susceptibility components may be related to each other through the reciprocity conditions~\eqref{eq:reciprocity} (see Appendix~\ref{app:recip}), which reduce the total number of independent unknowns. Such a consideration is not (yet) possible for the case of second-order nonlinear susceptibility tensors since there are currently no reciprocity relations available that would apply to these nonlinear susceptibility tensors, as discussed in Appendix~\ref{app:recip}.

However, it must be emphasized that a scattering particle generally only exhibits a few relevant nonlinear susceptibility components while most of the other terms are zero or at least may be neglected~\cite{LIU201853}. One of the reason for this is that a scattering particle, that may be practically realized to operate at optical frequencies, must necessarily have a relatively simple geometry due to the limitations of conventional fabrication techniques. Therefore, one may reduce the 108 nonlinear susceptibility components down to just a few relevant terms. For instance, when the scattering particle is symmetric in the metasurface plane, then its response to $x$- or $y$-polarized waves will be the same. In this case, the susceptibility tensors will exhibit further fundamental symmetries in addition to the intrinsic permutation symmetry consider above, hence greatly reducing the number of independent susceptibility unknowns.

Unfortunately, even if the number of unknowns may be reduced, the system~\eqref{eq:homo} generally remains under-determined. To overcome this issue, we may exploit the same approach as that conventionally used for linear metasurfaces: perform several simulations, each time for different pump incidence angles. Indeed, for each set of incidence angles $(\theta_\text{i}, \phi_\text{i})$, the susceptibilities remain the same but the incident and scattered fields differ, thus increasing the number of linearly independent equations, while keeping the same number of unknowns. For example, if the total number of independent nonzero susceptibility components is 8, then the system~\eqref{eq:homo} becomes fully determined when only two sets of pump incidence angles are used.

Finally, we would like to mention that, besides the homogenization technique discussed here, which is essentially the nonlinear extension of the conventional method used to homogenize linear metasurfaces, there is another approach to retrieve the nonlinear susceptibilities. This alternative approach is discussed in~\cite{Koendrink2014Retrieval}. It consists in exciting the structure with contra-propagating waves so that their magnetic (electric) fields cancel at the position of the structure, while their electric (magnetic) constructively interfere. It is thus possible to selectively excite specific nonlinear susceptibilities, for instance only $\te{\chi}_\text{eee}$. Then, by measuring the scattered fields and by changing the parameters of the contra-propagating waves, one may completely characterize the metasurface susceptibilities.

\subsection{Scattering Analysis}
\label{sec:scattering}

When the metasurface effective linear and nonlinear susceptibilities are known, it is possible to predict how the metasurface will scatter light for any pump incidence angles. The general approach to compute the fields scattered by the metasurface is to solve the system~\eqref{eq:homo} for the unknown reflected and transmitted fields.

In what follows and for simplicity, we concentrate our attention on the particular case where the pump is normally incident and the metasurface is spatially uniform. In this case, the spatial derivatives in~\eqref{eq:freqGSCTs} are zero since both the fields and the susceptibilities are spatially uniform in the metasurface plane, which greatly simplifies the scattered field analysis. Indeed, since the spatial derivatives are zero, the presence of normal polarizations (and hence susceptibilities) may be ignored since they do not contribute to the scattering. This means that the only relevant components of the incident and scattered electromagnetic fields, $E_i$ and $H_i$ in~\eqref{eq:homo}, as well as the linear $(\chi^{ij})$ and nonlinear $(\chi^{ijk})$ susceptibility tensors are such that $i,j,k = \{x,y\}$.

This approach allows us to rigorously compute the fields scattered by the metasurface. However, in many cases, the incident wave is not normally impinging on the metasurface and/or the latter may be spatially varying. In these scenarios, the spatial derivatives in~\eqref{eq:freqGSCTs} do not vanish. Nevertheless, if the incidence angle is small and/or the spatial variations of the susceptibilities are slow compared to the wavelength at $2\omega$, then it is possible to neglect the presence of the spatial derivatives (small angle approximation). Accordingly, it is possible to obtain an approximate evaluation of the scattering from the metasurface, which gets worse as the incidence angle and/or spatial variations increase~\cite{Achouri2016e}.

To compute the second-harmonic scattering from nonlinear metasurfaces, we simplify the GSTCs and write them in a more compact and convenient form. To do so, we start by expressing the magnetic field in terms of the corresponding electric field in the case of normally propagating plane waves. We thus only consider their transverse components. For plane waves propagating in the forward (+$z$) and backward (-$z$) directions, we respectively have that
\begin{equation}
\label{eq:EH}
\ve{H}_\text{fw} = \frac{1}{\eta_0}\te{\text{J}}\cdot\ve{E}_\text{fw}, \quad
\ve{H}_\text{bw} = -\frac{1}{\eta_0}\te{\text{J}}\cdot\ve{E}_\text{bw},
\end{equation}
where $\te{\text{J}}$ is the rotation matrix defined as
\begin{equation}
\te{\text{J}}=
\begin{pmatrix}
0&&-1\\
1&&0
\end{pmatrix}.
\end{equation}
Then, by making use of~\eqref{eq:EH}, we express the average of the pump electromagnetic field that is used in~\eqref{eq:NLPM}. When the pump propagates normally in the forward direction, this average field is given by
\begin{subequations}
\label{eq:avFW}
\begin{equation}
\label{eq:avFW1}
\ve{E}_\text{av,fw}^{\omega} = \frac{1}{2}\left(\te{\text{I}} + \te{S}_{11} + \te{S}_{21} \right)\cdot\ve{E}_0^\omega,
\end{equation}
\begin{equation}
\label{eq:avFW2}
\ve{H}_\text{av,fw}^{\omega} = \frac{1}{2\eta_0}\te{\text{J}}\cdot\left(\te{\text{I}} - \te{S}_{11} + \te{S}_{21} \right)\cdot\ve{E}_0^\omega,
\end{equation}
\end{subequations}
where $\ve{E}_0^\omega$ is the amplitude of the pump electric field, the matrices $\te{S}$ are the linear scattering matrices obtained in Appendix~\ref{app:Sparam}, $\te{\text{I}}$ is the two-dimensional identity matrix. Note that ports 1 and 2 respectively refer to the left ($z<0$) and right ($z>0$) sides of the metasurface.

Similarly, if the pump is propagating in the backward direction, then the average pump fields are
\begin{subequations}
\label{eq:avBW}
\begin{equation}
\label{eq:avBW1}
\ve{E}_\text{av,bw}^{\omega} = \frac{1}{2}\left(\te{\text{I}} + \te{S}_{22} + \te{S}_{12} \right)\cdot\ve{E}_0^\omega,
\end{equation}
\begin{equation}
\label{eq:avBW2}
\ve{H}_\text{av,bw}^{\omega} = -\frac{1}{2\eta_0}\te{\text{J}}\cdot\left(\te{\text{I}} - \te{S}_{22} + \te{S}_{12} \right)\cdot\ve{E}_0^\omega.
\end{equation}
\end{subequations}

We now express the averages of the second-harmonic fields that are used in~\eqref{eq:PM}. Since there is no wave impinging on the metasurface at $2\omega$, these average fields are simply given by
\begin{subequations}
\label{eq:LavFW}
\begin{equation}
\ve{E}_\text{av}^{2\omega} = \frac{1}{2}\left(\ve{E}_\text{fw}^{2\omega} + \ve{E}_\text{bw}^{2\omega}\right),
\end{equation}
\begin{equation}
\ve{H}_\text{av}^{2\omega} = \frac{1}{2\eta_0}\te{\text{J}}\cdot\left(\ve{E}_\text{fw}^{2\omega} - \ve{E}_\text{bw}^{2\omega}\right).
\end{equation}
\end{subequations}
By the same token, the difference of the fields in~\eqref{eq:freqGSCTs} read
\begin{subequations}
\label{eq:LdiffFW}
\begin{equation}
\Delta\ve{E}^{2\omega} = \frac{1}{2}\left(\ve{E}_\text{fw}^{2\omega} - \ve{E}_\text{bw}^{2\omega}\right),
\end{equation}
\begin{equation}
\Delta\ve{H}^{2\omega} = \frac{1}{2\eta_0}\te{\text{J}}\cdot\left(\ve{E}_\text{fw}^{2\omega} + \ve{E}_\text{bw}^{2\omega}\right).
\end{equation}
\end{subequations}
The GSTCs may now be simplified by substituting~\eqref{eq:LavFW} and~\eqref{eq:LdiffFW}, along with~\eqref{eq:SplitPM} and~\eqref{eq:PM}, into~\eqref{eq:freqGSCTs}. We thus get
\begin{subequations}
\begin{equation}
\begin{split}
\ve{E}_\text{fw}^{2\omega} + \ve{E}_\text{bw}^{2\omega} &= -j\frac{\omega}{c_0}\Big[\te{\chi}_\text{ee}\cdot(\ve{E}_\text{fw}^{2\omega}+\ve{E}_\text{bw}^{2\omega})\\
&+\te{\chi}_\text{em}\cdot\te{\text{J}}\cdot(\ve{E}_\text{fw}^{2\omega}-\ve{E}_\text{bw}^{2\omega}) \Big] - j \omega\eta_0\ve{P}_\text{nl}^{2\omega},
\end{split}
\end{equation}
\begin{equation}
\begin{split}
\te{\text{J}}\cdot(\ve{E}_\text{fw}^{2\omega} - \ve{E}_\text{bw}^{2\omega}) &= -j\frac{\omega}{c_0}\Big[\te{\chi}_\text{mm}\cdot\te{\text{J}}\cdot(\ve{E}_\text{fw}^{2\omega}-\ve{E}_\text{bw}^{2\omega})\\
&+\te{\chi}_\text{me}\cdot(\ve{E}_\text{fw}^{2\omega}+\ve{E}_\text{bw}^{2\omega}) \Big] - j \omega\mu_0\ve{M}_\text{nl}^{2\omega},
\end{split}
\end{equation}
\end{subequations}
where $c_0$ is the speed of light in vacuum.

These two vectorial equations may now be solved so as to express the backward and forward second-harmonic waves as functions of the linear susceptibilities and the nonlinear polarizations. They thus respectively read
\begin{subequations}
\label{eq:NLscat}
\begin{equation}
\ve{E}_\text{bw}^{2\omega} = \frac{1}{\epsilon_0}\te{\text{C}}_1 \cdot\ve{P}_\text{nl}^{2\omega} +\eta_0\te{\text{C}}_2 \cdot\ve{M}_\text{nl}^{2\omega},
\end{equation}
\begin{equation}
\ve{E}_\text{fw}^{2\omega} = \frac{1}{\epsilon_0}\te{\text{C}}_3 \cdot\ve{P}_\text{nl}^{2\omega} -\eta_0\te{\text{C}}_4 \cdot\ve{M}_\text{nl}^{2\omega},
\end{equation}
\end{subequations}
where the matrices $\te{\text{C}}$, which contain the linear susceptibilities, are explicitly provided in Appendix~\ref{app:param}. Relations~\eqref{eq:NLscat} are the general expressions for the second-harmonic fields scattered by a nonlinear metasurface. In these expressions, the nonlinear polarizations, $\ve{P}_\text{nl}^{2\omega}$ and $\ve{M}_\text{nl}^{2\omega}$, play the role of nonlinear sources excited by the pump at $\omega$. These polarizations are given in~\eqref{eq:NLPM} and are different for different directions of pump propagation according to~\eqref{eq:EH}.\\

From~\eqref{eq:NLscat}, we see that, as it is the case for linear metasurfaces, the forward or backward scattering can be completely suppressed since the terms on the right-hand sides may cancel each other if the linear and nonlinear susceptibilities are properly engineered. It should be therefore possible to obtain the second-order nonlinear counterparts of the Kerker conditions~\cite{Kerker83}. However, in the very general formulation~\eqref{eq:NLscat}, it is difficult to obtain the nonlinear reflectionless conditions solely in terms of the susceptibilities because the nonlinear polarizations densities depend upon the scattering of the pump fields, as can be seen in~\eqref{eq:avFW} and~\eqref{eq:avBW}. Indeed, it is, in general, particularly cumbersome to factor the linear scattering matrices out of the double dot products in~\eqref{eq:NLPM}, since~\eqref{eq:avFW2} (or~\eqref{eq:avBW2}) cannot be expressed in terms of~\eqref{eq:avFW1} (or~\eqref{eq:avBW1}).

Nevertheless, this issue may be overcome by considering a specific but particularly interesting and insightful situation. The case where the metasurface is reflectionless for the pump excitation. This means that, at frequency $\omega$, the reflectionless conditions~\eqref{eq:linearRLC} in Appendix~\ref{app:RLC} are satisfied and, hence, that $\te{S}_{11}=\te{S}_{22}= 0$ in~\eqref{eq:avFW} and~\eqref{eq:avBW}. In that case, the average forward and backward magnetic fields,~\eqref{eq:avFW2} and~\eqref{eq:avBW2}, are, using~\eqref{eq:EH}, simply given by
\begin{equation}
\label{eq:EHav}
\ve{H}_\text{av,fw}^\omega = \frac{1}{\eta_0}\te{\text{J}}\cdot\ve{E}_\text{av,fw}^\omega, \quad
\ve{H}_\text{av,bw}^\omega = -\frac{1}{\eta_0}\te{\text{J}}\cdot\ve{E}_\text{av,bw}^\omega.
\end{equation}

Based on the assumption that the pump excitation is not reflected, the nonlinear polarization densities in~\eqref{eq:NLPM} may be simplified using~\eqref{eq:EHav}. For a forward propagating pump, they become\footnote{Where we have used the fact that $\ve{E}\ve{H} = \ve{E}\cdot\ve{H}^T ~\propto~  \ve{E}\cdot(\te{\text{J}}\cdot\ve{E})^T = \ve{E}\cdot\ve{E}^T\cdot\te{\text{J}}^T = -\ve{E}\ve{E}\cdot\te{\text{J}}$, where $T$ is the transpose operation and $\te{\text{J}}^T = -\te{\text{J}}$. Similarly $\ve{H}\ve{H}~\propto~ -\te{\text{J}}\cdot\ve{E}\ve{E}\cdot\te{\text{J}}$.}
\begin{subequations}
\label{eq:FWPM}
\begin{equation}
\begin{split}
&\ve{P}^{2\omega}_\text{nl,fw} =\frac{\epsilon_0}{2}\Big(\te{\chi}_\text{eee}:\ve{E}_\text{av,fw}^\omega\ve{E}_\text{av,fw}^\omega \\
&- \te{\chi}_\text{eem}:\ve{E}_\text{av,fw}^\omega\ve{E}_\text{av,fw}^\omega\cdot\te{\text{J}} - \te{\chi}_\text{emm}:\te{\text{J}}\cdot\ve{E}_\text{av,fw}^\omega\ve{E}_\text{av,fw}^\omega\cdot\te{\text{J}}\Big),
\end{split}
\end{equation}
\begin{equation}
\begin{split}
&\ve{M}^{2\omega}_\text{nl,fw}=\frac{1}{2\eta_0}\Big(-\te{\chi}_\text{mmm}:\te{\text{J}}\cdot\ve{E}_\text{av,fw}^\omega\ve{E}_\text{av,fw}^\omega\cdot\te{\text{J}}
\\
&\qquad-\te{\chi}_\text{mem}:\ve{E}_\text{av,fw}^\omega\ve{E}_\text{av,fw}^\omega\cdot\te{\text{J}} + \te{\chi}_\text{mee}:\ve{E}_\text{av,fw}^\omega\ve{E}_\text{av,fw}^\omega\Big),
\end{split}
\end{equation}
\end{subequations}
and for a backward propagating pump, they become
\begin{subequations}
\label{eq:BWPM}
\begin{equation}
\begin{split}
&\ve{P}^{2\omega}_\text{nl,bw} =\frac{\epsilon_0}{2}\Big(\te{\chi}_\text{eee}:\ve{E}_\text{av,bw}^\omega\ve{E}_\text{av,bw}^\omega + \te{\chi}_\text{eem}:\ve{E}_\text{av,bw}^\omega\ve{E}_\text{av,bw}^\omega\cdot\te{\text{J}}\\
&\quad - \te{\chi}_\text{emm}:\te{\text{J}}\cdot\ve{E}_\text{av,bw}^\omega\ve{E}_\text{av,bw}^\omega\cdot\te{\text{J}}\Big),
\end{split}
\end{equation}
\begin{equation}
\begin{split}
&\ve{M}^{2\omega}_\text{nl,bw}=\frac{1}{2\eta_0}\Big(-\te{\chi}_\text{mmm}:\te{\text{J}}\cdot\ve{E}_\text{av,bw}^\omega\ve{E}_\text{av,bw}^\omega\cdot\te{\text{J}}\\
&\quad+\te{\chi}_\text{mem}:\ve{E}_\text{av,bw}^\omega\ve{E}_\text{av,bw}^\omega\cdot\te{\text{J}} + \te{\chi}_\text{mee}:\ve{E}_\text{av,bw}^\omega\ve{E}_\text{av,bw}^\omega\Big).
\end{split}
\end{equation}
\end{subequations}
We may now further simplify relations~\eqref{eq:FWPM} and~\eqref{eq:BWPM} by applying the rotation matrices, $\te{\text{J}}$, directly on the nonlinear susceptibility tensors instead of the electric field matrices, $\ve{E}_\text{av}^\omega\ve{E}_\text{av}^\omega$. These rotations of the susceptibility tensors are presented in Appendix~\ref{app:param}. With this simplification, it is possible to factor the $\ve{E}_\text{av}^\omega\ve{E}_\text{av}^\omega$ terms out of the polarization densities. Consequently, the second-harmonic scattered fields in~\eqref{eq:NLscat} may thus be expressed in the following compact form:
\begin{subequations}
\label{eq:scattering}
\begin{equation}
\ve{E}_\text{bw,fw}^{2\omega}=\te{S}_{11}^{\omega\rightarrow 2\omega}:\ve{E}_\text{av}^\omega\ve{E}_\text{av}^\omega,
\end{equation}
\begin{equation}
\ve{E}_\text{bw,bw}^{2\omega}=\te{S}_{22}^{\omega\rightarrow 2\omega}:\ve{E}_\text{av}^\omega\ve{E}_\text{av}^\omega,
\end{equation}
\begin{equation}
\ve{E}_\text{fw,fw}^{2\omega}=\te{S}_{21}^{\omega\rightarrow 2\omega}:\ve{E}_\text{av}^\omega\ve{E}_\text{av}^\omega,
\end{equation}
\begin{equation}
\ve{E}_\text{fw,bw}^{2\omega}=\te{S}_{12}^{\omega\rightarrow 2\omega}:\ve{E}_\text{av}^\omega\ve{E}_\text{av}^\omega,
\end{equation}
\end{subequations}
where the first subscripts of the terms on the left-hand sides correspond to the direction of propagation of the scattered fields, while the second subscripts correspond to the direction of propagation of the pump. In~\eqref{eq:scattering}, we have introduced the notion of nonlinear scattering tensors, which are defined as
\begin{subequations}
\label{eq:NLSP}
\begin{equation}
\begin{split}
\te{S}_{11}^{\omega\rightarrow 2\omega} &= \frac{1}{2}\te{\text{C}}_1\cdot\left(\te{\chi}_\text{eee} -\te{\chi}_\text{eem}' - \te{\chi}_\text{emm}' \right)\\
&\quad+\frac{1}{2}\te{\text{C}}_2\cdot\left(-\te{\chi}_\text{mmm}' -\te{\chi}_\text{mem}' + \te{\chi}_\text{mee}' \right),
\end{split}
\end{equation}
\begin{equation}
\begin{split}
\te{S}_{22}^{\omega\rightarrow 2\omega} &= \frac{1}{2}\te{\text{C}}_3\cdot\left(\te{\chi}_\text{eee} +\te{\chi}_\text{eem}' - \te{\chi}_\text{emm}' \right)\\
&\quad+\frac{1}{2}\te{\text{C}}_4\cdot\left(\te{\chi}_\text{mmm}' -\te{\chi}_\text{mem}' - \te{\chi}_\text{mee}' \right).
\end{split}
\end{equation}
\begin{equation}
\begin{split}
\te{S}_{21}^{\omega\rightarrow 2\omega} &= \frac{1}{2}\te{\text{C}}_3\cdot\left(\te{\chi}_\text{eee} -\te{\chi}_\text{eem}' - \te{\chi}_\text{emm}' \right)\\
&\quad+\frac{1}{2}\te{\text{C}}_4\cdot\left(\te{\chi}_\text{mmm}' +\te{\chi}_\text{mem}' - \te{\chi}_\text{mee}' \right),
\end{split}
\end{equation}
\begin{equation}
\begin{split}
\te{S}_{12}^{\omega\rightarrow 2\omega} &= \frac{1}{2}\te{\text{C}}_1\cdot\left(\te{\chi}_\text{eee} +\te{\chi}_\text{eem}' - \te{\chi}_\text{emm}' \right)\\
&\quad+\frac{1}{2}\te{\text{C}}_2\cdot\left(-\te{\chi}_\text{mmm}' +\te{\chi}_\text{mem}' + \te{\chi}_\text{mee}' \right),
\end{split}
\end{equation}
\end{subequations}
where the primed tensors are those that have been rotated according to the operations provided in Appendix~\ref{app:param}.

At this point, it is important to realize that the nonlinear scattering tensors defined in~\eqref{eq:NLSP} are third-order tensors and not just simple matrices like the conventional linear scattering tensors used in Appendix~\ref{app:Sparam} and in Eqs.~\eqref{eq:avFW} and~\eqref{eq:avBW}. Accordingly, there is a total number of 32 scattering parameters in~\eqref{eq:NLSP}, while there is only 16 linear scattering parameters in~\eqref{eq:Smatrix} in Appendix~\ref{app:Sparam}.

In the absence of external time-odd bias, the metasurface is linearly reciprocal (refer to Appendix~\ref{app:recip}), which is the case of the vast majority of metasurfaces. In that case, relations~\eqref{eq:SPreciprocity} apply. Therefore, the 16 linear scattering parameters in~\eqref{eq:Smatrix} are reduced to only 10 independent parameters.

\begin{table*}[t]
  \centering
\begin{tabular}{c|c|c|c}
\hline
 & \begin{tabular}[c]{@{}c@{}}Absence of bianisotropy\\ $\te{\chi}_\text{em} = \te{\chi}_\text{me}=0$\\~\end{tabular} & \begin{tabular}[c]{@{}c@{}}Linearly reflectionless\\ $\te{\chi}_\text{ee} = -\te{\text{J}}\cdot\te{\chi}_\text{mm}\cdot\te{\text{J}}$\\ $\te{\chi}_\text{em} = \te{\text{J}}\cdot\te{\chi}_\text{me}\cdot\te{\text{J}}$\end{tabular} & \begin{tabular}[c]{@{}c@{}}Combination of both\\ $\te{\chi}_\text{ee} = -\te{\text{J}}\cdot\te{\chi}_\text{mm}\cdot\te{\text{J}}$\\ $\te{\chi}_\text{em} = \te{\chi}_\text{me}=0$\end{tabular} \\ \hline
   \begin{tabular}[c]{@{}c@{}}
Parameters\\in Appendix~\ref{app:param}
\end{tabular}                                       & \begin{tabular}[c]{@{}c@{}} $\te{\text{C}}_1=\te{\text{C}}_3=-j\left(\frac{c_0}{\omega}\te{\text{I}}+j\te{\chi}_\text{ee}\right)^{-1}$\\
 $\te{\text{C}}_2=\te{\text{C}}_4=-j\left(\frac{c_0}{\omega}\te{\text{J}}+j\te{\chi}_\text{mm}\cdot\te{\text{J}}\right)^{-1}\cdot\te{\text{J}}$   \end{tabular}                                                                          & \begin{tabular}[c]{@{}c@{}} $\te{\text{C}}_1=\te{\text{C}}_2=-j\left(\frac{c_0}{\omega}\te{\text{I}}+j\te{\chi}_\text{ee}-j\te{\chi}_\text{em}\cdot\te{\text{J}}\right)^{-1}$\\
 $\te{\text{C}}_3=\te{\text{C}}_4=-j\left(\frac{c_0}{\omega}\te{\text{I}}+j\te{\chi}_\text{ee}+j\te{\chi}_\text{em}\cdot\te{\text{J}}\right)^{-1}$   \end{tabular}                                                                                            &  \begin{tabular}[c]{@{}c@{}}  $\te{\text{C}}_1=\te{\text{C}}_2=\te{\text{C}}_3=\te{\text{C}}_4$\\
 $=-j\left(\frac{c_0}{\omega}\te{\text{I}}+j\te{\chi}_\text{ee}\right)^{-1}$  \end{tabular}     \\ \hline
$\te{S}_{11}^{\omega \rightarrow 2\omega}=0$                         &    \begin{tabular}[c]{@{}c@{}}
$\te{\text{C}}_1\cdot\left(\te{\chi}_\text{eee} -\te{\chi}_\text{eem}' - \te{\chi}_\text{emm}' \right)=$\\
$\te{\text{C}}_2\cdot\left(\te{\chi}_\text{mmm}' +\te{\chi}_\text{mem}' - \te{\chi}_\text{mee}' \right)$
\end{tabular}                                                          & \multicolumn{2}{c}{     \begin{tabular}[c]{@{}c@{}}
$\te{\chi}_\text{eee} -\te{\chi}_\text{eem}' - \te{\chi}_\text{emm}' =$
$\te{\chi}_\text{mmm}' +\te{\chi}_\text{mem}' - \te{\chi}_\text{mee}'$
\end{tabular}   }                                                                              \\\hline
$\te{S}_{22}^{\omega \rightarrow 2\omega}=0$     &    \begin{tabular}[c]{@{}c@{}}
$\te{\text{C}}_1\cdot\left(\te{\chi}_\text{eee} +\te{\chi}_\text{eem}' - \te{\chi}_\text{emm}' \right)=$\\
$-\te{\text{C}}_2\cdot\left(\te{\chi}_\text{mmm}' -\te{\chi}_\text{mem}' - \te{\chi}_\text{mee}' \right)$
\end{tabular}    &    \multicolumn{2}{c}{ \begin{tabular}[c]{@{}c@{}}
$\te{\chi}_\text{eee} +\te{\chi}_\text{eem}' - \te{\chi}_\text{emm}'=$
$-\te{\chi}_\text{mmm}' +\te{\chi}_\text{mem}' + \te{\chi}_\text{mee}'$
\end{tabular}    }                                                        \\ \hline
$\te{S}_{21}^{\omega \rightarrow 2\omega}=0$  & \begin{tabular}[c]{@{}c@{}}
$\te{\text{C}}_1\cdot\left(\te{\chi}_\text{eee} -\te{\chi}_\text{eem}' - \te{\chi}_\text{emm}' \right)=$\\
$-\te{\text{C}}_2\cdot\left(\te{\chi}_\text{mmm}' +\te{\chi}_\text{mem}' - \te{\chi}_\text{mee}' \right)$
\end{tabular}   &   \multicolumn{2}{c}{
$\te{\chi}_\text{eee} -\te{\chi}_\text{eem}' - \te{\chi}_\text{emm}'=$
$-\te{\chi}_\text{mmm}' -\te{\chi}_\text{mem}' +\te{\chi}_\text{mee}' $ }     \\ \hline
$\te{S}_{12}^{\omega \rightarrow 2\omega}=0$  & \begin{tabular}[c]{@{}c@{}}
$\te{\text{C}}_1\cdot\left(\te{\chi}_\text{eee} +\te{\chi}_\text{eem}' - \te{\chi}_\text{emm}' \right)=$\\
$\te{\text{C}}_2\cdot\left(\te{\chi}_\text{mmm}' -\te{\chi}_\text{mem}' - \te{\chi}_\text{mee}' \right)$
\end{tabular}   & \multicolumn{2}{c}{
$\te{\chi}_\text{eee} +\te{\chi}_\text{eem}' - \te{\chi}_\text{emm}' =$
$\te{\chi}_\text{mmm}' -\te{\chi}_\text{mem}' - \te{\chi}_\text{mee}' $ }   \\ \hline
\begin{tabular}[c]{@{}c@{}}Asym. reflection\\ $\te{S}_{11}^{\omega \rightarrow 2\omega} \neq \te{S}_{22}^{\omega \rightarrow 2\omega}$\end{tabular} & $-\te{\text{C}}_1\cdot\te{\chi}_\text{eem}' -\te{\text{C}}_2\cdot\left(\te{\chi}_\text{mmm}'-\te{\chi}_\text{mee}'\right)\neq0$ & \begin{tabular}[c]{@{}c@{}}$\frac{1}{2}\te{\text{C}}_1\cdot\Big(\te{\chi}_\text{eee} -\te{\chi}_\text{eem}' - \te{\chi}_\text{emm}' -\te{\chi}_\text{mmm}' $\\$-\te{\chi}_\text{mem}' + \te{\chi}_\text{mee}' \Big)-\frac{1}{2}\te{\text{C}}_3\cdot\Big(\te{\chi}_\text{eee} +\te{\chi}_\text{eem}'$\\$ - \te{\chi}_\text{emm}' +\te{\chi}_\text{mmm}' -\te{\chi}_\text{mem}' - \te{\chi}_\text{mee}' \Big)\neq0$ \end{tabular}& $-\te{\text{C}}_1\cdot\left(\te{\chi}_\text{eem}' +\te{\chi}_\text{mmm}'-\te{\chi}_\text{mee}'\right)\neq0$                                                          \\ \hline
\begin{tabular}[c]{@{}c@{}}Asym. transmission\\ $\te{S}_{21}^{\omega \rightarrow 2\omega} \neq \te{S}_{12}^{\omega \rightarrow 2\omega}$\end{tabular}& $-\te{\text{C}}_1\cdot\te{\chi}_\text{eem}' +\te{\text{C}}_2\cdot\left(\te{\chi}_\text{mmm}'-\te{\chi}_\text{mee}'\right)\neq0$ & \begin{tabular}[c]{@{}c@{}}$\frac{1}{2}\te{\text{C}}_3\cdot\Big(\te{\chi}_\text{eee} -\te{\chi}_\text{eem}' - \te{\chi}_\text{emm}' +\te{\chi}_\text{mmm}' $\\$+\te{\chi}_\text{mem}' - \te{\chi}_\text{mee}' \Big)-\frac{1}{2}\te{\text{C}}_1\cdot\Big(\te{\chi}_\text{eee} +\te{\chi}_\text{eem}'$\\$ - \te{\chi}_\text{emm}' -\te{\chi}_\text{mmm}' +\te{\chi}_\text{mem}' + \te{\chi}_\text{mee}' \Big)\neq0$ \end{tabular} & $-\te{\text{C}}_1\cdot\left(\te{\chi}_\text{eem}' -\te{\chi}_\text{mmm}'+\te{\chi}_\text{mee}'\right)\neq0$                                                           \\ \hline
\end{tabular}
  \caption{Nonlinear reflectionless and transmissionless conditions as well as the susceptibilities responsible for asymmetric reflection and transmission for three different scenarios. 1) In the absence of bianisotropy, 2) when the linear reflectionless conditions are satisfied, and 3) when both conditions are simultaneously satisfied.}
  \label{tab:1}
\end{table*}

According to the discussion in Appendix~\ref{app:recip}, the nonlinear scattering tensors~\eqref{eq:NLSP} are not subjected to any reciprocal condition. However, they are affected by the intrinsic permutation symmetries of the structure. This means that $S^{ijk}_\text{ab}=S^{ikj}_\text{ab}$, where $i,j,k = \{x,y\}$ and $\text{a,b} = \{1,2\}$. Consequently, the 32 scattering parameters in~\eqref{eq:NLSP} are reduced to 24 independent parameters. Hence, a second-order nonlinear metasurface exhibits much more degrees of freedom available to control the scattered fields compared to conventional linear metasurfaces. More specifically, a nonlinear metasurface has the particularly interesting property of exhibiting asymmetric second-harmonic generation. For instance, a nonlinear metasurface may be perfectly \emph{reciprocal} and have\footnote{In~\cite{valev2014nonlinear,poutrina2014multipole,Poutrina2016,Achouri2017NLMS}, it was claimed that this inequality represents a nonreciprocal operation. However, according to the upcoming discussion in this section as well as that in Appendix~\ref{app:recip}, we shall rather refer to it as an asymmetric rather than a nonreciprocal operation.} $S_{21}^{\omega\rightarrow 2\omega, xxx}\neq S_{12}^{\omega\rightarrow 2\omega, xxx}$, while, by reciprocity, in the linear regime the equality $S_{21}^{xx}= S_{12}^{xx}$ must be satisfied according to~\eqref{eq:SPreciprocity}.

\begin{figure}[h!]
\centering
\subfloat[]{\label{fig:NLMSexample1}
\includegraphics[width=0.5\columnwidth]{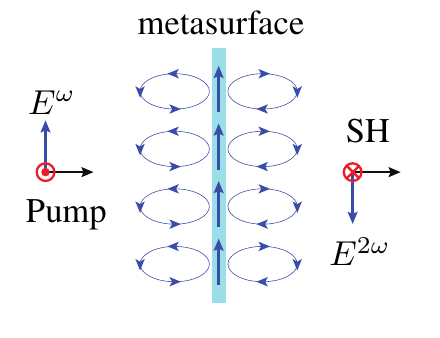}
}
\subfloat[]{\label{fig:NLMSexample2}
\includegraphics[width=0.5\columnwidth]{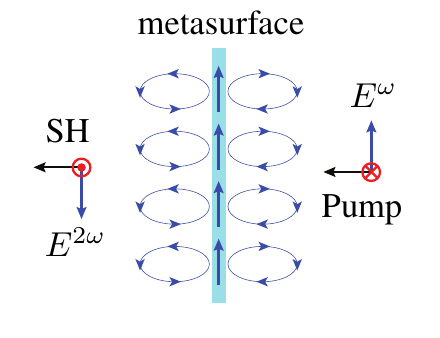}
}\\
\subfloat[]{\label{fig:NLMSexample3}
\includegraphics[width=0.5\columnwidth]{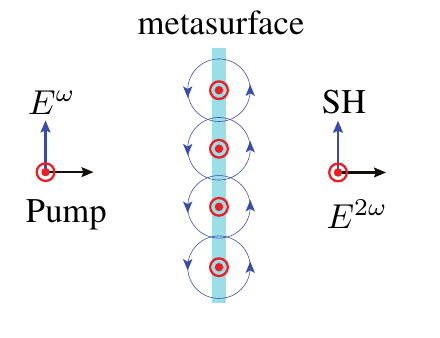}
}
\subfloat[]{\label{fig:NLMSexample4}
\includegraphics[width=0.5\columnwidth]{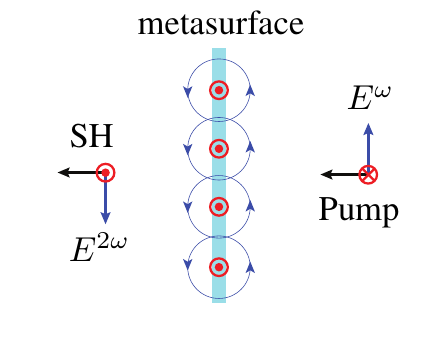}
}
\caption{Comparison of second-harmonic scattering from two different nonlinear metasurfaces. In (a) and (b), the metasurface is electrically nonlinear and nonlinear electric dipolar moments are excited within the metasurface. In (c) and (d), it is magnetically nonlinear and nonlinear magnetic dipolar moments are excited within the metasurface. In (a) and (c), the pump propagates forward, while in (b) and (d), it propagates backward. In all figures, the blue arrows around the dipolar moments represent the corresponding scattered electric fields.}
\label{fig:NLMSexample}
\end{figure}

In order to understand why, for a reciprocal metasurface, $S_{21}^{\omega\rightarrow 2\omega, xxx}$ can be different from $ S_{12}^{\omega\rightarrow 2\omega, xxx}$, let us consider the two following idealized situations. In the first situation, a pump at frequency $\omega$ illuminates a nonlinear metasurface which only presents $\chi_\text{eee}^{xxx}$ as a nonzero susceptibility, while all other susceptibility terms are assumed to be zero or negligible. In the second situation, the nonlinear metasurface only presents $\chi_\text{mmm}^{yyy}$ as a nonzero susceptibility. For simplicity, we consider that\footnote{Note that we are using \emph{surface} susceptibilities, hence the dimension of linear susceptibilities is [m] instead of being dimensionless like it is the case for bulk susceptibilities. Moreover, the nonlinear electric and magnetic susceptibilities have the same dimension thanks to the convention that we have used in~\eqref{eq:NLPM}.} $\chi_\text{eee}^{xxx} = \chi_\text{mmm}^{yyy} = 1~\text{m}^2/\text{V}$.

These two metasurfaces are successively excited with a pump propagating once in the forward direction and then in the backward direction. In both cases, the pump excites the metasurface susceptibilities and hence the nonlinear polarizations, which are responsible for the second-harmonic generation.

The scattering (here, for simplicity, second-harmonic transmitted field only) from the electrically nonlinear metasurface is depicted in Figs.~\ref{fig:NLMSexample1} and~\ref{fig:NLMSexample2} for the two excitation directions. Similarly, the scattering from the magnetically nonlinear metasurface is depicted in Figs.~\ref{fig:NLMSexample3} and~\ref{fig:NLMSexample4}.

From these representations, it is clear that the electrically nonlinear metasurface produces the same transmitted field irrespectively of the direction of pump propagation since $P_x^{2\omega}=\chi_\text{eee}^{xxx}E_x^2$ is symmetric with respect to $E_x$, thus $S_{21}^{\omega \rightarrow 2\omega} = S_{12}^{\omega \rightarrow 2\omega}$. However, the opposite occurs in the case of the magnetically nonlinear metasurface since $M_y^{2\omega}=\chi_\text{mmm}^{yyy}H_y^2$ is antisymmetric with respect to $H_y$ when the direction of propagation is reversed and thus $S_{21}^{\omega \rightarrow 2\omega} \neq S_{12}^{\omega \rightarrow 2\omega}$. This simple example shows that the presence of this magnetic nonlinear susceptibility introduces an \emph{asymmetric} second-harmonic generation in transmission.

In what follows, we will consider the nonlinear scattering tensors in~\eqref{eq:NLSP} and generalize the concept of asymmetric nonlinear scattering. We will also look into other important aspects of nonlinear metasurfaces, which notably includes the conditions that enable one to completely suppress the forward or backward second-harmonic generation.

Table~\ref{tab:1} summarizes the upcoming results. We consider three different scenarios, which are reported in the columns of the table. In order of appearance, we consider: 1) the absence of bianisotropy such that $\te{\chi}_\text{em} = \te{\chi}_\text{me}=0$ at all frequencies, which is usually the case if the surface is symmetric in the longitudinal ($z$) direction and if it exhibits no chirality (the scattered fields, at $\omega$, have the same polarization as the excitation, also at $\omega$); 2) that the linear reflectionless conditions in~\eqref{eq:linearRLC} are satisfied at both $\omega$ and $2\omega$; and 3) that both conditions 1) and 2) are simultaneously satisfied. In the corresponding rows of the table, we start by providing the updated $\te{\text{C}}$ tensors for the three respective scenarios. Then, we present the general reflectionless and transmissionless conditions for both forward and backward pump propagations. Finally, we provide the properties of asymmetric reflection and transmission.

Being able to suppress either the reflected or transmitted fields is achieved by superposition of the fields scattered by both electric and magnetic dipolar moments. By controlling the phase-shift between these two dipoles, it is thus possible to completely cancel the field scattered either in the backward or the forward direction. In the case of linear structures, this effect is referred to as the Kerker condition~\cite{Kerker83} and has been extensively used to realize reflectionless (linear) metasurfaces. The case of nonlinear metasurfaces is fundamentally identical when nonlinear electric and magnetic dipoles can be excited. For instance, we see that an electrically ($\te{\chi}_\text{eee} \neq 0$) and magnetically ($\te{\chi}_\text{mmm} \neq 0$) nonlinear metasurface, which satisfies the linear reflectionless conditions, is nonlinearly reflectionless for a forward propagating pump ($\te{S}_{11}^{\omega \rightarrow 2\omega}=0$) when $\te{\chi}_\text{eee}=\te{\chi}_\text{mmm}'$, which corresponds to the nonlinear counterpart of~\eqref{eq:linearRLC1}. However, for a backward propagating pump, the corresponding reflectionless condition ($\te{S}_{22}^{\omega \rightarrow 2\omega}=0$) is $\te{\chi}_\text{eee}=-\te{\chi}_\text{mmm}'$. The fact that the reflectionless conditions are not the same from both sides is another evidence of the asymmetric second-harmonic scattering behavior of these types of nonlinear metasurfaces. In fact, the two last rows of the table provide the expressions of the susceptibility components respectively responsible for the metasurface asymmetric reflection ($\te{S}_{11}^{\omega \rightarrow 2\omega}\neq \te{S}_{22}^{\omega \rightarrow 2\omega}$) and transmission ($\te{S}_{21}^{\omega \rightarrow 2\omega} \neq \te{S}_{12}^{\omega \rightarrow 2\omega}$). In the absence of bianisotropy, we see that the susceptibility tensors $\te{\chi}_\text{eem}'$, $\te{\chi}_\text{mmm}'$ and $\te{\chi}_\text{mee}'$ are inducing some sorts of scattering asymmetry, which is a generalization of the concept already illustrated in Fig.~\ref{fig:NLMSexample}. However, in the presence of bianisotropy, all nonlinear susceptibilities are naturally inducing asymmetric scattering since, to achieve bianisotropy, the metasurface has to be spatially asymmetric, as discussed before.

It is also interesting to note that the reflectionless and transmissionless conditions provided in Table~\ref{tab:1} are satisfied when the terms on the left-hand sides are equal to the terms on the right-hand sides of the equalities, since electric and magnetic terms should cancel each other. Therefore, in the absence of magnetic nonlinear susceptibilities, if the electric susceptibilities $\te{\chi}_\text{eee}$, $\te{\chi}_\text{eem}'$ and $\te{\chi}_\text{emm}'$ cancel out then the nonlinear scattering is completely suppressed both in reflection and transmission.

\section{Time-Domain Approach}
\label{Sec:time}

We shall now discuss the approach which consists in solving~\eqref{eq:GSTCs} in the time domain. This method is of practical interest when the depletion of the pump must be considered. This approach is generally mathematically more involved than the frequency-domain technique. Indeed, for the latter, it is relatively simple to obtain the second-order scattering response of the metasurface since we assume that the pump is undepleted. In contrast, the time-domain approach naturally takes into account the pump depletion as well as the generation of higher-order harmonic due to the interactions of the signals at $\omega$ and $2\omega$ (and so on) with the nonlinear metasurface. Moreover, in the time-domain formulation, the dispersive nature of the susceptibilities is conventionally expressed as time convolutions with the acting field such that the polarization densities in~\eqref{eq:GSTCs} take the following general form~\cite{rothwell2008electromagnetics}:
\begin{equation}
\label{eq:dispersion}
\ve{P}(\ve{r},t) = \epsilon_0\int_{-\infty}^{t} dt' ~ \te{\chi}_\text{ee}(\ve{r},t-t')\cdot\ve{E}(\ve{r},t) +\ldots
\end{equation}

In order to overcome these additional difficulties, we next assume that the susceptibilities are dispersionless. Although this seems a rather stringent assumption, we shall remember that the field interacting with the metasurface may generally be expressed as $\ve{E}(\ve{r},\omega) = \sum_{n=1}^{\infty} \ve{E}_n(\ve{r}) e^{jn\omega t}$. Moreover, the most important terms of this sum are typically those for which $n=1$ (linear) and $n=2$ (second harmonic). Considering~\eqref{eq:dispersion}, this means that in reality the susceptibilities must satisfy the condition $\chi(\omega)=\chi(2\omega)$, which corresponds to the conventional nonlinear phase matching condition~\cite{boyd2003nonlinear}, rather than being dispersionless at any frequency.

In what follows, we will solve~\eqref{eq:GSTCs} in the time-domain and discuss both the operations of homogenization and scattered field analysis. However, we will not do it in a general fashion, as we did for the frequency-domain method, because of the complexity of the time-domain approach. We shall rather illustrate the method to solve~\eqref{eq:GSTCs} with an example similar to that used in~\cite{Achouri2017NLMS}. Accordingly, let us consider the case of an electrically and magnetically nonlinear metasurface that only exhibits the following nonzero susceptibility components: $\chi_\text{ee}^{xx}, \chi_\text{mm}^{yy}, \chi_\text{eee}^{xxx}$ and $\chi_\text{mmm}^{yyy}$. In that case, the time-domain GSTCs reduce to
\begin{subequations}
\label{eq:TDgstc}
\begin{equation}
-\Delta H=\epsilon_0\chi_\text{ee}^{xx}\frac{\partial}{\partial t}E_\text{av} +\epsilon_0\chi_\text{eee}^{xxx}\frac{\partial}{\partial t}E_\text{av}^2,
\end{equation}
\begin{equation}
-\Delta E=\mu_0\chi_\text{mm}^{yy}\frac{\partial}{\partial t}H_\text{av} +\mu_0\eta_0\chi_\text{mmm}^{yyy}\frac{\partial}{\partial t}H_\text{av}^2,
\end{equation}
\end{subequations}
where we assume that the waves are $x$-polarized and normally propagating. For a forward propagating pump and by making use of~\eqref{eq:EH}, we have that
\begin{subequations}
\label{eq:TDgstcSimple}
\begin{equation}
\begin{split}
-E_\text{fw}-&E_\text{bw}+E_\text{pump}=\frac{\eta_0\epsilon_0}{2}\chi_\text{ee}^{xx}\frac{\partial}{\partial t}\left(E_\text{fw}+E_\text{bw}+E_\text{pump}\right) \\
&+\frac{\eta_0\epsilon_0}{4}\chi_\text{eee}^{xxx}\frac{\partial}{\partial t}\left(E_\text{fw}+E_\text{bw}+E_\text{pump}\right)^2,
\end{split}
\end{equation}
\begin{equation}
\begin{split}
-E_\text{fw}+&E_\text{bw}+E_\text{pump}=\frac{\mu_0}{2\eta_0}\chi_\text{mm}^{yy}\frac{\partial}{\partial t}\left(E_\text{fw}-E_\text{bw}+E_\text{pump}\right) \\
& +\frac{\mu_0}{2\eta_0}\chi_\text{mmm}^{yyy}\frac{\partial}{\partial t}\left(E_\text{fw}-E_\text{bw}+E_\text{pump}\right)^2.
\end{split}
\end{equation}
\end{subequations}
This system of equation may now be solved either to obtain the susceptibilities in terms of known fields or to get the scattered fields in terms of known susceptibilities. In~\cite{Achouri2017NLMS}, we already provide the susceptibilities in terms of known fields, therefore we do not present them here again. However, the scattering from such a metasurface was only presented for the particular case where the linear and nonlinear reflectionless conditions are satisfied, thus only the transmission coefficients were provided. We shall now address the more general situation where the linear and nonlinear reflectionless are not necessarily satisfied leading to both reflected and transmitted fields.

As it is, the system~\eqref{eq:TDgstcSimple} forms a set of two first-order inhomogeneous coupled nonlinear differential equations, which does not possess analytical solutions. It is however possible to obtain approximate expressions of the scattered fields using perturbation theory. We assume that the forward and backward scattered fields may be respectively expressed as
\begin{subequations}
\label{eq:PTe}
\begin{equation}
E_\text{fw} \approx E_\text{0,fw} + \gamma E_\text{1,fw} + \gamma^2 E_\text{2,fw} + \ldots + \gamma^n E_\text{n,fw},
\end{equation}
\begin{equation}
E_\text{bw} \approx E_\text{0,bw} + \gamma E_\text{1,bw} + \gamma^2 E_\text{2,bw} + \ldots + \gamma^n E_\text{n,bw},
\end{equation}
\end{subequations}
where $\gamma$ is a small quantity. We also consider the following conditions on the susceptibilities:
\begin{equation}
\label{eq:PTx}
\chi_\text{ee}^{xx}\gg \chi_\text{eee}^{xxx} \sim \gamma,~\text{and}~\chi_\text{mm}^{yy}\gg \chi_\text{mmm}^{yyy} \sim \gamma.
\end{equation}

These conditions reflect the difference in terms of amplitude between linear and nonlinear susceptibilities. For instance, in conventional optical systems the value of gamma is about $\gamma \sim 10^{-12}$~\cite{boyd2003nonlinear}.

It is now possible to obtain the approximate expression of the backward, $E_\text{bw}$, and forward, $E_\text{fw}$, scattered fields by inserting~\eqref{eq:PTe} and~\eqref{eq:PTx} into~\eqref{eq:TDgstcSimple} and using $E_\text{pump} = E_0 \cos{(\omega t)}$.

Now, for a given value of $n$, the $n$-th term of expressions~\eqref{eq:PTe} can be solved for by removing all terms proportional to $\gamma^{m}$ with $m > n$, which greatly simplifies the complexity of the system~\eqref{eq:TDgstcSimple}. By doing so, we derive the first 4 terms of the backward and forward scattered fields expansions. It turns out that each of these terms contains a certain number of harmonics such that the n-th term is proportional to the following harmonic(s):
\begin{subequations}
\begin{align}
\label{eq:order}
&n=0 \rightarrow \omega,\\
&n=1 \rightarrow 2\omega,\\
&n=2 \rightarrow \omega, 3\omega,\\
&n=3 \rightarrow 2\omega, 4\omega.
\end{align}
\end{subequations}

Finally, the fields scattered at frequency $\omega$ may be found by combining the contributions from the $n=0$ and $n=2$ terms, while the fields scattered at frequency $2\omega$ may be found using the $n=1$ and $n=3$ terms, and so on. The resulting fields scattered at $\omega$ are\footnote{In these expressions, as well as in~\eqref{eq:2wTDscattered}, the wavenumber is at frequency $\omega$, i.e. $k=\omega/c_0$.}
\begin{subequations}
\label{eq:1wTDscattered}
\begin{equation}
\begin{split}
&E_\text{fw}^{\omega} = E_0\frac{4+\chi_\text{ee}^{xx}\chi_\text{mm}^{yy}k^2}{(2+jk\chi_\text{ee}^{xx})(2+jk\chi_\text{mm}^{yy})}\\
&-4E_0^3k^2\Bigg( \frac{(\chi_\text{eee}^{xxx})^2}{(1+jk\chi_\text{ee}^{xx})(2-jk\chi_\text{ee}^{xx})(2+jk\chi_\text{ee}^{xx})^3}\\
&\quad\qquad+\frac{(\chi_\text{mmm}^{yyy})^2}{(1+jk\chi_\text{mm}^{yy})(2-jk\chi_\text{mm}^{yy})(2+jk\chi_\text{mm}^{yy})^3} \Bigg),
\end{split}
\end{equation}
\begin{equation}
\begin{split}
&E_\text{bw}^{\omega} = E_0\frac{2jk(\chi_\text{mm}^{yy}-\chi_\text{ee}^{xx})}{(2+jk\chi_\text{ee}^{xx})(2+jk\chi_\text{mm}^{yy})}\\
&-4E_0^3k^2\Bigg( \frac{(\chi_\text{eee}^{xxx})^2}{(1+jk\chi_\text{ee}^{xx})(2-jk\chi_\text{ee}^{xx})(2+jk\chi_\text{ee}^{xx})^3}\\
&\quad\qquad-\frac{(\chi_\text{mmm}^{yyy})^2}{(1+jk\chi_\text{mm}^{yy})(2-jk\chi_\text{mm}^{yy})(2+jk\chi_\text{mm}^{yy})^3} \Bigg),
\end{split}
\end{equation}
\end{subequations}
where the first terms (proportional to $E_0$) on the right-hand sides correspond to the undepleted pump approximations. In fact, these two terms are the solutions that are found for the scattering of a conventional linear metasurface~\cite{achouri2014general} and which may be derived directly from~\eqref{eq:reducedSysInv}. The second terms (proportional to $E_0^3$) on the right-hand sides are correction terms that model the depletion of the pump. Note that these correction terms are directly proportional to the square of the wavenumber and of the nonlinear susceptibilities, this means that they are generally weak compared to the linear contributions except for very large values of $E_0$.

We next present the solutions corresponding to the scattered fields at $2\omega$. However, due to the length of these expressions, we only provide the $n=1$ terms, which read
\begin{subequations}
\label{eq:2wTDscattered}
\begin{equation}
\begin{split}
E_\text{fw}^{2\omega} = -2jE_0^2k  &\Bigg(\frac{\chi_\text{eee}^{xxx}}{(1+jk\chi_\text{ee}^{xx})(2+jk\chi_\text{ee}^{xx})^2}\\
&+ \frac{\chi_\text{mmm}^{yyy}}{(1+jk\chi_\text{mm}^{yy})(2+jk\chi_\text{mm}^{yy})^2}\Bigg),
\end{split}
\end{equation}
\begin{equation}
\begin{split}
E_\text{bw}^{2\omega} = -2jE_0^2k  &\Bigg(\frac{\chi_\text{eee}^{xxx}}{(1+jk\chi_\text{ee}^{xx})(2+jk\chi_\text{ee}^{xx})^2}\\
&- \frac{\chi_\text{mmm}^{yyy}}{(1+jk\chi_\text{mm}^{yy})(2+jk\chi_\text{mm}^{yy})^2}\Bigg).
\end{split}
\end{equation}
\end{subequations}
Again, these terms correspond to the undepleted pump approximation. Adding the $2\omega$ contributions from the $n=3$ terms would provide a correction which takes into account the depletion of the pump. As a consequence, the results in~\eqref{eq:2wTDscattered} are \emph{exactly} those that would be obtained using the general frequency-domain scattering relations~\eqref{eq:NLscat} with the assumption that $\chi(\omega)=\chi(2\omega)$.

The time-domain approach presented here is thus more complicated to use than the frequency-domain one discussed before. However, the main advantage of this time-domain approach is that it allows one to take into account the depletion of the pump, which may be useful when the second-harmonic conversion efficiency of nonlinear metasurfaces will become significant. Finally, we also mention the fact that the time-domain technique may also be converted into a finite-difference time-domain technique, as discussed in~\cite{Achouri2017NLMS}.

\section{Conclusion}
\label{Sec:conc}

In this work, we have presented an elaborate discussion on the electromagnetic theory of second-harmonic generation in nonlinear metasurfaces. We have focused our attention on the homogenization and second-harmonic scattering analysis of such structures. Both a frequency-domain and a time-domain approach have been presented.

It is clear that with the current conversion efficiency of nonlinear metasurfaces, the frequency-domain approach, which assumes an undepleted pump regime, is the most convenient of the two techniques.

Moreover, we have tried to remain as general as possible so as to cover all possible scenarios. Accordingly, we have derived the general reflectionless and transmissionless conditions and highlighted the fundamental reasons of asymmetric reflection and transmission in these structures. We have also clarified the concept of asymmetric scattering versus nonreciprocal scattering in nonlinear media and when changes in frequency are considered.

\appendices

\section{Homogenization and Scattering of\\ Linear Metasurfaces}
\label{app:Sparam}

In this appendix, we briefly present the main steps required to homogenize and obtain the scattering parameters of linear bianisotropic metasurfaces~\cite{achouri2017design}. We will here assume that these metasurfaces are uniform and that the incident wave propagates normally so that the spatial derivatives in~\eqref{eq:GSTCs} may be dropped\footnote{The very general case of oblique incidence may be treated following the exact same procedure as that described in Sec.~\ref{sec:homotech}}. In the case of a bianisotropic linear metasurface, the GSTCs read~\cite{achouri2014general}
\begin{subequations}
\label{eq:InvProb}
\begin{align}
\ve{\hat{z}}\times\Delta\ve{H}
&=j\omega\epsilon_0\te{\chi}_\text{ee}\cdot\ve{E}_\text{av}+jk_0\te{\chi}_\text{em}\cdot\ve{H}_\text{av},\label{eq:diffH}\\
\Delta\ve{E}\times\ve{\hat{z}}
&=j\omega\mu_0 \te{\chi}_\text{mm}\cdot\ve{H}_\text{av}+jk_0\te{\chi}_\text{me}\cdot\ve{E}_\text{av}.\label{eq:diffE}
\end{align}
\end{subequations}
It is often particularly convenient to cast this system of equations into a matrix form to simplify the upcoming computations. Accordingly, the system becomes
\begin{equation}
\label{eq:InvProbMatrix}
\begin{pmatrix}
\Delta H_y\\
\Delta H_x\\
\Delta E_y\\
\Delta E_x
\end{pmatrix}=
\begin{pmatrix}
\widetilde{\chi}_\text{ee}^{xx} & \widetilde{\chi}_\text{ee}^{xy} & \widetilde{\chi}_\text{em}^{xx} & \widetilde{\chi}_\text{em}^{xy}\\
\widetilde{\chi}_\text{ee}^{yx} & \widetilde{\chi}_\text{ee}^{yy} & \widetilde{\chi}_\text{em}^{yx} & \widetilde{\chi}_\text{em}^{yy}\\
\widetilde{\chi}_\text{me}^{xx} & \widetilde{\chi}_\text{me}^{xy} & \widetilde{\chi}_\text{mm}^{xx} & \widetilde{\chi}_\text{mm}^{xy}\\
\widetilde{\chi}_\text{me}^{yx} & \widetilde{\chi}_\text{me}^{yy} & \widetilde{\chi}_\text{mm}^{yx} & \widetilde{\chi}_\text{mm}^{yy}
\end{pmatrix}
\cdot
\begin{pmatrix}
E_{x,\text{av}}\\
E_{y,\text{av}}\\
H_{x,\text{av}}\\
H_{y,\text{av}}
\end{pmatrix},
\end{equation}
where the relationship between the susceptibilities in~\eqref{eq:InvProb} and the normalized susceptibilities in~\eqref{eq:InvProbMatrix}, is given by
\begin{equation}
\label{eq:conv}
\begin{split}
&
\begin{pmatrix}
\chi_\text{ee}^{xx} & \chi_\text{ee}^{xy} & \chi_\text{em}^{xx} & \chi_\text{em}^{xy}\\
\chi_\text{ee}^{yx} & \chi_\text{ee}^{yy} & \chi_\text{em}^{yx} & \chi_\text{em}^{yy}\\
\chi_\text{me}^{xx} & \chi_\text{me}^{xy} & \chi_\text{mm}^{xx} & \chi_\text{mm}^{xy}\\
\chi_\text{me}^{yx} & \chi_\text{me}^{yy} & \chi_\text{mm}^{yx} & \chi_\text{mm}^{yy}
\end{pmatrix}=\\
&\quad
=\begin{pmatrix}
\frac{j}{\omega\epsilon_0}\widetilde{\chi}_\text{ee}^{xx} & \frac{j}{\omega\epsilon_0}\widetilde{\chi}_\text{ee}^{xy} & \frac{j}{k_0}\widetilde{\chi}_\text{em}^{xx} & \frac{j}{k_0}\widetilde{\chi}_\text{em}^{xy}\\
-\frac{j}{\omega\epsilon_0}\widetilde{\chi}_\text{ee}^{yx} & -\frac{j}{\omega\epsilon_0}\widetilde{\chi}_\text{ee}^{yy} & -\frac{j}{k_0}\widetilde{\chi}_\text{em}^{yx} & -\frac{j}{k_0}\widetilde{\chi}_\text{em}^{yy}\\
-\frac{j}{k_0}\widetilde{\chi}_\text{me}^{xx} & -\frac{j}{k_0}\widetilde{\chi}_\text{me}^{xy} & -\frac{j}{\omega\mu_0}\widetilde{\chi}_\text{mm}^{xx} & -\frac{j}{\omega\mu_0}\widetilde{\chi}_\text{mm}^{xy}\\
\frac{j}{k_0}\widetilde{\chi}_\text{me}^{yx} & \frac{j}{k_0}\widetilde{\chi}_\text{me}^{yy} & \frac{j}{\omega\mu_0}\widetilde{\chi}_\text{mm}^{yx} & \frac{j}{\omega\mu_0}\widetilde{\chi}_\text{mm}^{yy}
\end{pmatrix}.
\end{split}
\end{equation}

We now write the matrix system~\eqref{eq:InvProbMatrix} in the following compact form:
\begin{equation}
\label{eq:reducedSys}
 \te{\Delta} = \widetilde{\te{\chi}}\cdot \te{A}_v,
\end{equation}
where $\te{\Delta}$, $\widetilde{\te{\chi}}$ and $\te{A}_v$ refer to the field differences, the normalized susceptibilities and the field averages, respectively.

From this system of equations we can now easily homogenize the metasurface. This may be done by illuminating the metasurface with a normally incident plane wave and computing the electric and magnetic scattered fields. From these known fields, we can define the components of the field differences, $\te{\Delta}$, and averages, $\te{A}_v$, and ultimately obtain the susceptibilities, $\widetilde{\te{\chi}}$, by matrix inversion of~\eqref{eq:reducedSys}. However, this system of equations contains 16 unknown susceptibilities for only 4 equations, and is thus under-determined. In order to solve it, we consider 4 different illuminations instead of just 1. As a consequence, the system now contains 16 equations (4 for each illumination) for the same 16 unknown susceptibilities, and is now fully determined. The 4 illuminations that we consider are: forward $x$-polarization, forward $y$-polarization, backward $x$-polarization and backward $y$-polarization.

For each of these illuminations and resulting scattered fields, we express the corresponding electric and magnetic fields in terms of scattering parameters. For instance, the incident, reflected and transmitted electric fields, in the case of a forward $x$-polarized excitation, are respectively given by
\begin{equation}
\label{eq:xPol}
\ve{E}_{\text{i}}=\ve{\hat{x}},
\quad
\ve{E}_{\text{r}}=S_{11}^{xx}\ve{\hat{x}} + S_{11}^{yx}\ve{\hat{y}},
\quad
\ve{E}_{\text{t}}=S_{21}^{xx}\ve{\hat{x}} + S_{21}^{yx}\ve{\hat{y}},
\end{equation}
where $S_{ab}^{uv}$, with $a, b = \{1,2\}$ and $u, v = \{x,y\}$, are the scattering parameters. We consider that port 1 is on the left ($z<0$) of the metasurface, while port 2 is on its right ($z>0$). We also consider that the metasurface is surrounded on both sides by different media with intrinsic impedance $\eta_1$ and $\eta_2$, respectively.

Expressing the electromagnetic fields of the 4 illuminations in the same fashion as in~\eqref{eq:xPol}, leads, after simplification, to the matrices $\te{\Delta}$ and $\te{A}_v$ given below (see Eqs.~\eqref{eq:DelAv}), where the matrices $\te{S}_{ab}$ and $\te{N}$ are defined by
\begin{subequations}
\label{eq:DelAv}
\begin{floatEq}
\begin{equation}
\label{eq:deltaMat}
\te{\Delta}=
\begin{pmatrix}
-\te{N}/\eta_1 + \te{N}\cdot\te{S}_{11}/\eta_1 + \te{N}\cdot\te{S}_{21}/\eta_2 & -\te{N}/\eta_2 +\te{N}\cdot\te{S}_{12}/\eta_1 + \te{N}\cdot\te{S}_{22}/\eta_2 \\
-\te{\text{J}}\cdot\te{N} -\te{\text{J}}\cdot\te{N}\cdot\te{S}_{11} +\te{\text{J}}\cdot\te{N}\cdot\te{S}_{21} &\te{\text{J}}\cdot\te{N} -\te{\text{J}}\cdot\te{N}\cdot\te{S}_{12}+\te{\text{J}}\cdot\te{N}\cdot\te{S}_{22}
\end{pmatrix},
\end{equation}
\begin{equation}
\label{eq:AvMat}
\te{A}_v=\frac{1}{2}
\begin{pmatrix}
\te{I} + \te{S}_{11}+ \te{S}_{21} &
\te{I} + \te{S}_{12}+ \te{S}_{22}
\\
\te{\text{J}}/\eta_1 -\te{\text{J}}\cdot\te{S}_{11}/\eta_1 +\te{\text{J}}\cdot\te{S}_{21}/\eta_2 &
-\te{\text{J}}/\eta_2 -\te{\text{J}}\cdot\te{S}_{12}/\eta_1 +\te{\text{J}}\cdot\te{S}_{22}/\eta_2
\end{pmatrix}.
\end{equation}
\end{floatEq}
\end{subequations}
\begin{equation}
\label{eq:defS}
\te{S}_{ab}=
\begin{pmatrix}
S_{ab}^{xx} & S_{ab}^{xy} \\
S_{ab}^{yx} & S_{ab}^{yy}
\end{pmatrix},\qquad
\te{N}=
\begin{pmatrix}
1 & 0 \\
0 & -1
\end{pmatrix}.
\end{equation}

Instead of homogenizing a metasurface, we may use the system~\eqref{eq:reducedSys} to find to fields scattered by a metasurface with known susceptibilities. To do this, we insert~\eqref{eq:DelAv} into~\eqref{eq:reducedSys} and solve for the scattering parameters. We thus obtain the following relation:
\begin{equation}
\label{eq:reducedSysInv}
 \te{S} = \te{M}_1^{-1}\cdot\te{M}_2,
\end{equation}
where $\te{S}$ is a $4\times 4$ matrix defined as
\begin{equation}\label{eq:Smatrix}
\te{S}=
\begin{pmatrix}
\te{S}_{11} & \te{S}_{12} \\
\te{S}_{21} & \te{S}_{22}
\end{pmatrix},
\end{equation}
and the resulting matrices $\te{M}_1$ and $\te{M}_2$ are given below (see Eqs.~\eqref{eq:Mat12}).
\begin{subequations}
\label{eq:Mat12}
\begin{floatEq}
\begin{equation}
\label{eq:Mat1}
\te{M}_1=
\begin{pmatrix}
\te{N}/\eta_1 - \widetilde{\te{\chi}}_\text{ee}/2 + \widetilde{\te{\chi}}_\text{em}\cdot\te{\text{J}}/(2\eta_1) & \te{N}/\eta_2 - \widetilde{\te{\chi}}_\text{ee}/2 - \widetilde{\te{\chi}}_\text{em}\cdot\te{\text{J}}/(2\eta_2) \\
-\te{\text{J}}\cdot\te{N} - \widetilde{\te{\chi}}_\text{me}/2 + \widetilde{\te{\chi}}_\text{mm}\cdot\te{\text{J}}/(2\eta_1) &\te{\text{J}}\cdot\te{N} - \widetilde{\te{\chi}}_\text{me}/2 - \widetilde{\te{\chi}}_\text{mm}\cdot\te{\text{J}}/(2\eta_2)
\end{pmatrix},
\end{equation}
\begin{equation}
\label{eq:Mat2}
\te{M}_2=
\begin{pmatrix}
\widetilde{\te{\chi}}_\text{ee}/2 + \te{N}/\eta_1+\widetilde{\te{\chi}}_\text{em}\cdot\te{\text{J}}/(2 \eta_1) & \widetilde{\te{\chi}}_\text{ee}/2 + \te{N}/\eta_2-\widetilde{\te{\chi}}_\text{em}\cdot\te{\text{J}}/(2 \eta_2) \\
\widetilde{\te{\chi}}_\text{me}/2 +\te{\text{J}}\cdot\te{N}+\widetilde{\te{\chi}}_\text{mm}\cdot\te{\text{J}}/(2 \eta_1) & \widetilde{\te{\chi}}_\text{me}/2 -\te{\text{J}}\cdot\te{N}-\widetilde{\te{\chi}}_\text{mm}\cdot\te{\text{J}}/(2 \eta_2)
\end{pmatrix}.
\end{equation}
\end{floatEq}
\end{subequations}

\section{Discussion on the nonreciprocity of metasurfaces}
\label{app:recip}

Nonreciprocity is a theoretically and practically important effect that is often misunderstood~\cite{Jalas2013WhatisIsolator}. In what follows, we provide a brief discussion on the concept of nonreciprocity which applies to linear and nonlinear media so as to clarify several points brought up in this paper.

In the case of a linear and time-invariant (LTI) system (a metasurface in our case), nonreciprocity may be achieved by breaking the time-reversal symmetry of the system. According to the Onsager-Casimir principle~\cite{Onsager,Casimir}, which provides the time-symmetry relations of tensorial constitutive parameters, the action of breaking time-reversal symmetry may be realized by externally biasing the system with a time-odd quantity~\cite{jackson_classical_1999}. A well-known example is that of Faraday isolators, which are implemented by biasing a ferrite with a static magnetic field~\cite{nye1985physical}.

In the absence of external bias, the Onsager-Casimir symmetry relations reduce to the conventional reciprocity conditions provided by the Lorentz reciprocity theorem~\cite{rothwell2008electromagnetics}. In the case of a bianisotropic medium, they are given by
\begin{equation}
\label{eq:reciprocity}
\te{\chi}_\text{ee}^\text{T}=\te{\chi}_\text{ee},\qquad
\te{\chi}_\text{mm}^\text{T}=\te{\chi}_\text{mm},\qquad
\te{\chi}_\text{me}^\text{T}=-\te{\chi}_\text{em}.
\end{equation}
An LTI metasurface is thus reciprocal if these conditions are satisfied.

In order to assess the nonreciprocal response of a system, it is often particularly convenient to analyze its scattering parameters. In the case of a two-port system, the following reciprocity conditions apply
\begin{equation}
\label{eq:SPreciprocity}
\te{S}_{21}^\text{T}=\te{S}_{12},\qquad
\te{S}_{11}^\text{T}=\te{S}_{11},\qquad
\te{S}_{22}^\text{T}=\te{S}_{22},
\end{equation}
where the scattering matrices have the form of~\eqref{eq:defS}. These conditions may naturally be extended to the case of an N-port network~\cite{pozar2011microwave}. An important consideration is that an LTI system may exhibit \emph{asymmetric} scattering such that $\te{S}_{21}^\text{T}\neq\te{S}_{21}$, while still being perfectly reciprocal, $\te{S}_{21}^\text{T}=\te{S}_{12}$. In fact, a spatially asymmetric LTI system is \emph{always} reciprocal expect if it is externally biased with a time-odd quantity as mentioned above.

The case of nonlinear media is more complicated. First of all, there is no nonlinear counterpart to the Lorentz reciprocity theorem~\cite{potton2004reciprocity}. Furthermore, spatial asymmetry is sufficient to achieve nonreciprocity~\cite{Roy2010opticalDiode}. Finally, we should consider two different situations: when the frequency of the excitation is changed by the nonlinear process, as it is the case with second-harmonic generation, and when it is not changed, as it is the case in Kerr media~\cite{boyd2003nonlinear}. For that latter situation, it should be noted that in specific cases, such as the operation of four-wave mixing in third-order nonlinear media, it is possible to obtain nonlinear Onsager relations~\cite{potton2004reciprocity,HubnerTimeReversalNLoptic}. But this does not apply to the case of second-order media, which notably change the frequency of the excitation and which are the topic of this paper.

In what follows, we shall discuss the nonreciprocal behavior of second-order nonlinear media. We illustrate this discussion with a simplified example. Consider the following one-dimensional (with no spatial variations in the $x$- and $y$-directions, i.e. $\partial/\partial y=\partial/\partial x=0$) two-port problem consisting of a second-order nonlinear metasurface surrounded by port 1 and port 2 respectively placed on its left- and right-hand sides. The metasurface is illuminated from port 1 with a pump field, ${E}_1$, at frequency $\omega$. The resulting transmitted field, ${E}_2$, is then measured at port 2. In a very general scenario, the field ${E}_2$ is proportional to all multiples of the fundamental harmonic, $\omega$. Assuming that all waves are $x$-polarized, the field ${E}_2$ can thus be expressed as
\begin{equation}
\label{eq:NLgeneralScat}
{E}_2 = {S}_{21}^{\omega\rightarrow \omega}{E}_1 + {S}_{21}^{\omega\rightarrow 2\omega}{E}_1^2 + {S}_{21}^{\omega\rightarrow 3\omega}{E}_1^3 +\ldots,
\end{equation}
where the scattering parameters correspond to the complex amplitude of the corresponding harmonic received at port~2.

The time-reversed $(-t)$ operation of the process described in~\eqref{eq:NLgeneralScat}, would consists in reversing the direction of wave propagation such that the field ${E}_2$ (and all its harmonics) emerges from port 2 and transmits back through the metasurface to then be received at port 1~\cite{HubnerTimeReversalNLoptic}. By symmetry, the field received at port 1 is exactly the same as the original pump field ${E_1}$ of the direct time $(t)$ scenario. Only in this case, would the system be considered as reciprocal and we would have that
\begin{equation}
\label{eq:NLgeneralScat2}
{S}_{21}^{\omega\rightarrow \omega}={S}_{12}^{\omega\rightarrow \omega},~ {S}_{21}^{\omega\rightarrow 2\omega}= {S}_{12}^{2\omega\rightarrow \omega},~ {S}_{21}^{\omega\rightarrow 3\omega}={S}_{12}^{3\omega\rightarrow \omega},~ \ldots
\end{equation}

It is important to realize that this time-reversed operation is a purely mathematical concept, which is practically impossible to implement. Indeed, it would be impossible to generate all the harmonics constituting $E_2$ and reproduce the exact phase shift between them so that the field received at port 1 is the same as the original field $E_1$~\cite{HubnerTimeReversalNLoptic} . Therefore, the equalities in~\eqref{eq:NLgeneralScat2} are generally not satisfied and the reversed operation is thus nonreciprocal. Accordingly, we generally have that
\begin{equation}
\label{eq:NLgeneralScat3}
{S}_{21}^{\omega\rightarrow \omega}\neq{S}_{12}^{\omega\rightarrow \omega},~ {S}_{21}^{\omega\rightarrow 2\omega}\neq {S}_{12}^{2\omega\rightarrow \omega},~ {S}_{21}^{\omega\rightarrow 3\omega}\neq{S}_{12}^{3\omega\rightarrow \omega},~ \ldots
\end{equation}
Note that if the system is spatially symmetric (in the $z$-direction) and that the field $E_2$ is generated at port 2 such that it contains only the fundamental harmonic, then the equality ${S}_{21}^{\omega\rightarrow \omega}={S}_{12}^{\omega\rightarrow \omega}$ is respected provided that the conditions~\eqref{eq:reciprocity} are satisfied. However, if the field $E_2$ contains only the frequency $2\omega$, then the equality ${S}_{21}^{\omega\rightarrow 2\omega}= {S}_{12}^{2\omega\rightarrow \omega}$ is generally not satisfied (even if the conditions~\eqref{eq:reciprocity} are satisfied). Such a scenario is discussed in~\cite{ShalaevTimeVaryingNR}, for the case of time-varying metasurfaces, which behave in a very similar fashion as nonlinear metasurfaces.

Finally, we point out that a nonlinear metasurface may exhibit \emph{asymmetric} nonlinear scattering, which has to be clearly differentiated from the \emph{nonreciprocal} nonlinear scattering described by relations~\eqref{eq:NLgeneralScat3}. Asymmetric second-hamonic scattering is defined as ${S}_{21}^{\omega\rightarrow 2\omega}\neq {S}_{12}^{\omega\rightarrow 2\omega}$, which essentially means that the second-harmonic field scattered in transmission by the metasurface is not the same as that when the latter is spatially flipped on itself. While this effect has often been referred to as a nonreciprocal process in the literature~\cite{valev2014nonlinear,poutrina2014multipole,Poutrina2016,Achouri2017NLMS}, the fact that ${S}_{21}^{\omega\rightarrow 2\omega}\neq {S}_{12}^{\omega\rightarrow 2\omega}$ is not due to time-reversal symmetry breaking, like the inequalities in~\eqref{eq:NLgeneralScat3}, but rather due to the asymmetric scattering of electric and magnetic nonlinear dipolar moments, as shown in Fig.~\ref{fig:NLMSexample}. Accordingly, this effect should not be referred to as nonreciprocal but rather as asymmetric scattering, which does not make it less interesting or potentially useful.

\section{Reduced Tensor Parameters}
\label{app:param}

In this appendix, we first provide the reduced linear susceptibility matrix parameters that are used to define general second-harmonic scattered field relations~\eqref{eq:NLscat} as well as the nonlinear scattering tensors in~\eqref{eq:scattering}. These reduced matrices read
\begin{subequations}
\begin{equation}
\begin{split}
\te{\text{C}}_1 = & -j2\te{\text{A}}\cdot\left(\frac{c_0}{\omega}\te{\text{J}}+j\te{\chi}_\text{mm}\cdot\te{\text{J}}+j\te{\chi}_\text{me} \right)\\
&\quad\cdot\left(\frac{c_0}{\omega}\te{\text{I}}+j\te{\chi}_\text{ee}+j\te{\chi}_\text{em}\cdot\te{\text{J}} \right)^{-1},
\end{split}
\end{equation}
\begin{equation}
\te{\text{C}}_2 = -j2\te{\text{A}}\cdot\te{\text{J}},
\end{equation}
\begin{equation}
\begin{split}
\te{\text{C}}_3 = & j2\left(\frac{c_0}{\omega}\te{\text{I}}+j\te{\chi}_\text{ee}+j\te{\chi}_\text{em}\cdot\te{\text{J}} \right)^{-1}\cdot\Bigg[\left(\frac{c_0}{\omega}\te{\text{I}}+j\te{\chi}_\text{ee}-j\te{\chi}_\text{em}\cdot\te{\text{J}} \right)\\
&\quad\cdot\te{\text{A}}\cdot\left(\frac{c_0}{\omega}\te{\text{J}}+j\te{\chi}_\text{mm}\cdot\te{\text{J}}+j\te{\chi}_\text{me} \right)\\
&\qquad\cdot\left(\frac{c_0}{\omega}\te{\text{I}}+j\te{\chi}_\text{ee}+j\te{\chi}_\text{em}\cdot\te{\text{J}} \right)^{-1}-\te{\text{I}}\Bigg],
\end{split}
\end{equation}
\begin{equation}
\begin{split}
\te{\text{C}}_4 = & -j2\left(\frac{c_0}{\omega}\te{\text{I}}+j\te{\chi}_\text{ee}+j\te{\chi}_\text{em}\cdot\te{\text{J}} \right)^{-1}\\
&\quad\cdot\left(\frac{c_0}{\omega}\te{\text{I}}+j\te{\chi}_\text{ee}-j\te{\chi}_\text{em}\cdot\te{\text{J}} \right)\cdot\te{\text{A}}\cdot\te{\text{J}},
\end{split}
\end{equation}
\end{subequations}
where the matrix $\te{\text{A}}$ is given by
\begin{equation}
\begin{split}
&\te{\text{A}} = \Bigg[\left(\frac{c_0}{\omega}\te{\text{J}}+j\te{\chi}_\text{mm}\cdot\te{\text{J}}-j\te{\chi}_\text{me} \right)+\Big(\frac{c_0}{\omega}\te{\text{J}}+j\te{\chi}_\text{mm}\cdot\te{\text{J}}+j\te{\chi}_\text{me} \Big)\\
&\quad\cdot\left(\frac{c_0}{\omega}\te{\text{I}}+j\te{\chi}_\text{ee}+j\te{\chi}_\text{em}\cdot\te{\text{J}} \right)^{-1}\cdot\left(\frac{c_0}{\omega}\te{\text{I}}+j\te{\chi}_\text{ee}-j\te{\chi}_\text{em}\cdot\te{\text{J}} \right)\Bigg]^{-1}.
\end{split}
\end{equation}

Next, we provide the relationships between the rotated nonlinear susceptibility tensors and their original form. These rotations affect the inner matrices of these third order tensors such that
\begin{subequations}
\begin{equation}
\left (\te{\chi}_\text{eem}'\right )_x =-\left (\te{\chi}_\text{eem}\right )_x\cdot\te{\text{J}} \quad \left (\te{\chi}_\text{eem}'\right )_y =-\left (\te{\chi}_\text{eem}\right )_y\cdot\te{\text{J}}
\end{equation}
\begin{equation}
\left (\te{\chi}_\text{emm}'\right )_x =\te{\text{J}}\cdot\left (\te{\chi}_\text{emm}\right )_x\cdot\te{\text{J}} \quad \left (\te{\chi}_\text{emm}'\right )_y =\te{\text{J}}\cdot\left (\te{\chi}_\text{emm}\right )_y\cdot\te{\text{J}}
\end{equation}
\begin{equation}
\left (\te{\chi}_\text{mmm}'\right )_x =-\te{\text{J}}\cdot\left (\te{\chi}_\text{mmm}\right )_y\cdot\te{\text{J}} \quad \left (\te{\chi}_\text{mmm}'\right )_y =\te{\text{J}}\cdot\left (\te{\chi}_\text{mmm}\right )_x\cdot\te{\text{J}}
\end{equation}
\begin{equation}
\left (\te{\chi}_\text{mem}'\right )_x =\left (\te{\chi}_\text{mem}\right )_y\cdot\te{\text{J}} \quad \left (\te{\chi}_\text{mem}'\right )_y =-\left (\te{\chi}_\text{mem}\right )_x\cdot\te{\text{J}}
\end{equation}
\begin{equation}
\left (\te{\chi}_\text{mee}'\right )_x =-\left (\te{\chi}_\text{mee}\right )_y \quad \left (\te{\chi}_\text{mee}'\right )_y =\left (\te{\chi}_\text{mee}\right )_x
\end{equation}
\end{subequations}

\section{Reflectionless Conditions for Linear Bianisotropic Metasurfaces}
\label{app:RLC}

As discussed in Sec.~\ref{sec:scattering}, the nonlinear reflectionless conditions directly depend on both the linear and nonlinear metasurface susceptibility tensors. Interestingly, it was shown that these \emph{nonlinear} reflectionless conditions are greatly simplified when the \emph{linear} reflectionless conditions are satisfied. We thus provide here the general reflectionless conditions for linear metasurfaces in the case of a normally incident plane wave.

These linear reflectionless conditions are obtained in a similar fashion as their nonlinear counter parts. The procedure to obtain them is to set $\te{S}_{11}=0$ and $\te{S}_{22}=0$ in~\eqref{eq:reducedSysInv} and solve for the susceptibilities. Accordingly, we get
\begin{subequations}
\label{eq:linearRLC}
\begin{equation}\label{eq:linearRLC1}
\te{\chi}_\text{ee} = -\te{\text{J}}\cdot\te{\chi}_\text{mm}\cdot\te{\text{J}},
\end{equation}
\begin{equation}
\label{eq:linearRLC2}
\te{\chi}_\text{em} = \te{\text{J}}\cdot\te{\chi}_\text{me}\cdot\te{\text{J}}.
\end{equation}
\end{subequations}
A particularly interesting scenario is when the metasurface is (linearly) reciprocal, which is most often the case. In this situation, the only way to simultaneously satisfy the condition~\eqref{eq:linearRLC2} and the reciprocity conditions~\eqref{eq:reciprocity} is when
\begin{equation}
\te{\chi}_\text{em} = \te{\chi}_\text{me} = \kappa\te{\text{I}},
\end{equation}
where $\kappa$ is a chiral coefficient. This equality implies that a reciprocal bianisotropic metasurface can only be reflectionless when it corresponds to a chiral bi-isotropic structure\cite{lindell1994electromagnetic}.

\bibliographystyle{myIEEEtran}
\bibliography{NewLib}

\end{document}